\documentclass[12pt]{iopart}

\newcommand{\be}{\begin{eqnarray}}
\newcommand{\ee}{\end{eqnarray}}
\newcommand{\ket}[1]{\ensuremath{\left| {#1} \right>}}
\newcommand{\bra}[1]{\ensuremath{\left< {#1} \right|}}
\newcommand{\braket}[2]{\ensuremath{\left<  {#1}\left|  {#2} \right.\right>}}

\newcommand{\caf}{\ensuremath{^{40}{\rm Ca}^{+}\, }}

\newcommand{\create}{\ensuremath{\,\hat{a}^{\dagger}}}
\newcommand{\destroy}{\ensuremath{\,\hat{a}}}

\usepackage{latexsym}
\usepackage{graphics}
\usepackage[squaren,thickspace,thickqspace]{SIunits}
\usepackage{multirow}
\usepackage{textcomp}

\begin{document}

\title[Quantum control of the motional states of trapped ions through fast switching]{Quantum control of the motional states of trapped ions through fast switching of trapping potentials}

\author{J~Alonso, F~M~Leupold, B~C~Keitch and J~P~Home}

\address{Institute for Quantum Electronics, ETH Z\"urich, Schafmattstrasse 16, 8093 Z\"urich, Switzerland}

\ead{alonso@phys.ethz.ch}

\begin{abstract}
We propose a new scheme for supplying voltages to the electrodes of microfabricated ion traps, enabling access to a regime in which changes to the trapping potential are made on timescales much shorter than the period of the secular oscillation frequencies of the trapped ions. This opens up possibilities for speeding up the transport of ions in segmented ion traps and also provides access to control of multiple ions in a string faster than the Coulomb interaction between them. We perform a theoretical study of ion transport using these methods in a surface-electrode trap, characterizing the precision required for a number of important control parameters. We also consider the possibilities and limitations for generating motional state squeezing using these techniques, which could be used as a basis for investigations of Gaussian-state entanglement.
\end{abstract}

\pacs{pacs}
\submitto{\NJP}
\maketitle

\section{Introduction}
\label{intro}

Engineered states of the motion of atomic ions in rf traps \cite{67Dehmelt,96Ghosh,05Major} have provided a number of pioneering demonstrations of quantum-state control and decoherence. Examples include the creation of Fock-state superpositions, squeezed states, and entangled states between the internal and motional degrees of freedom \cite{96Meekhof,96Monroe,00Myatt,07McDonnell,05Haljan,10Zahringer}. The motional degrees of freedom are also of primary importance in multi-qubit quantum logic gates, where the internal states can be entangled by transient state-dependent excitation of the motion \cite{03Leibfried,03SchmidtKaler}. More recently, motional degrees of freedom have been used as a conduit to transmit quantum states between different trap regions of a multi-zone trap array, whether by moving the ions themselves \cite{09Jost}, or by use of the long-range Coulomb interaction \cite{11Brown,11Harlander}.

The motion of trapped ions can be controlled by the application of optical or microwave fields, or by changing the electric fields which generate the trap itself. While optical and microwave fields can be used to create forces which are dependent on the ions' internal electronic state (and can therefore be used to produce spin-motion entanglement), this control can generally be approximated by a perturbative Hamiltonian with respect to that describing the trapping potential. By controlling the trap electric fields it is possible to produce large changes in the Hamiltonian describing the motion, and this control is independent of the ions' internal state.

Manipulating the motional states of trapped ions on fast timescales is desirable for a number of reasons. The primary application of current interest is quantum information processing, where transport of information is necessary for scaling to large numbers of ions, and has been predicted to be one of the most time-consuming operations in a large-scale processor \cite{06Schulz,03Steane}. One promising approach to scaling is to transport the ions themselves through an array of microtraps, with the transport controlled by time-varying potentials applied to a number of trap electrodes \cite{98Wineland,02Kielpinski}. These have traditionally been generated by voltage supplies placed outside the vacuum system and connected to the trap via vacuum feedthroughs. The feasibility of this approach has been demonstrated in a number of experiments, including linear transport (and separation) of multiple ions \cite{02Rowe,12Bowler,12Walther}, as well as 2-dimensional transport through T, X and Y junctions \cite{06Hensinger,10Amini,11Moehring,09Blakestad}. The important figures of merit in these experiments were the operation time and the amount of motional excitation which persists after the transport has taken place. The latter is important because errors in multi-qubit quantum logic gates increase for higher motional excitation. Though this can be mitigated by sympathetic cooling \cite{03Barrett, 09Home, 09Home2}, it takes time and thus reduces the computing speed of the processor \cite{10Hanneke}. As a result of these considerations, initial experiments on transport operated in the adiabatic regime, maintaining the ion close to the ground state of motion of the co-moving time-dependent potential well throughout the transport. The adiabatic constraint limits transport to timescales which are long compared to the motional period of oscillation of the ion in the trap.

The use of low-noise, high-speed Digital to Analog Converters (DACs) has recently enabled diabatic transport of ions over distances of $\sim\unit{300}{\micro\meter}$ in 5 to 16 oscillation cycles \cite{12Bowler,12Walther}, with the ion returned to the ground state of the potential well at the end of the transport despite being transiently excited during transport. With these techniques, the limit to the control rate is set by the finite capacitances present in the lines going to the electrodes.

In this paper, we propose time-dependent control of the trapping potential using a new method which involves placing electronic switches inside the vacuum system. The control for the switches is digital, allowing trap electrodes to be switched between two potentials supplied by standard analog supplies. Since the electronic switches we will consider can change their output by up to $\sim\unit{10}{\volt}$ on nanosecond timescales, this method provides the possibility of making changes to the trapping potentials 100 times faster than the period of oscillation of the secular motion, which is typically between \unit{200}{\nano\second} and \unit{1}{\micro\second} \cite{98Wineland}. This would allow for ion transport in times shorter than a single cycle of oscillation in the trap (section \ref{transport}).

Fast control of the trapping potential could also be used for other applications. In strings consisting of multiple ions, an additional important timescale is the speed of sound in the chain, which is characterized by the frequency splitting of normal modes. Controlling the potential on timescales fast compared to the speed of sound may allow for the realization of schemes which have been proposed for entanglement generation and the investigation of continuous-variable quantum information processing \cite{09Serafini,09Serafini2}.

This paper proceeds as follows. In section \ref{scheme} of the paper we introduce our fast control scheme. In section \ref{trap} we present a generic ion-trap setup which will be used as reference for the quantitative studies throughout the text. In section \ref{transport} we review the application of ultra-fast voltage switching to macroscopic ion transport in less than an axial oscillation cycle, while in section \ref{squeezing} we consider a different application to continuous quantum-variable manipulation. A number of further applications are outlined in section \ref{outlook}. Finally, a summary of the main results is given in section \ref{summary}.

\section{Electronic switches for ultra-fast motional control}
\label{scheme}

\begin{figure}
\centering
\resizebox{1\textwidth}{!}{
  \includegraphics{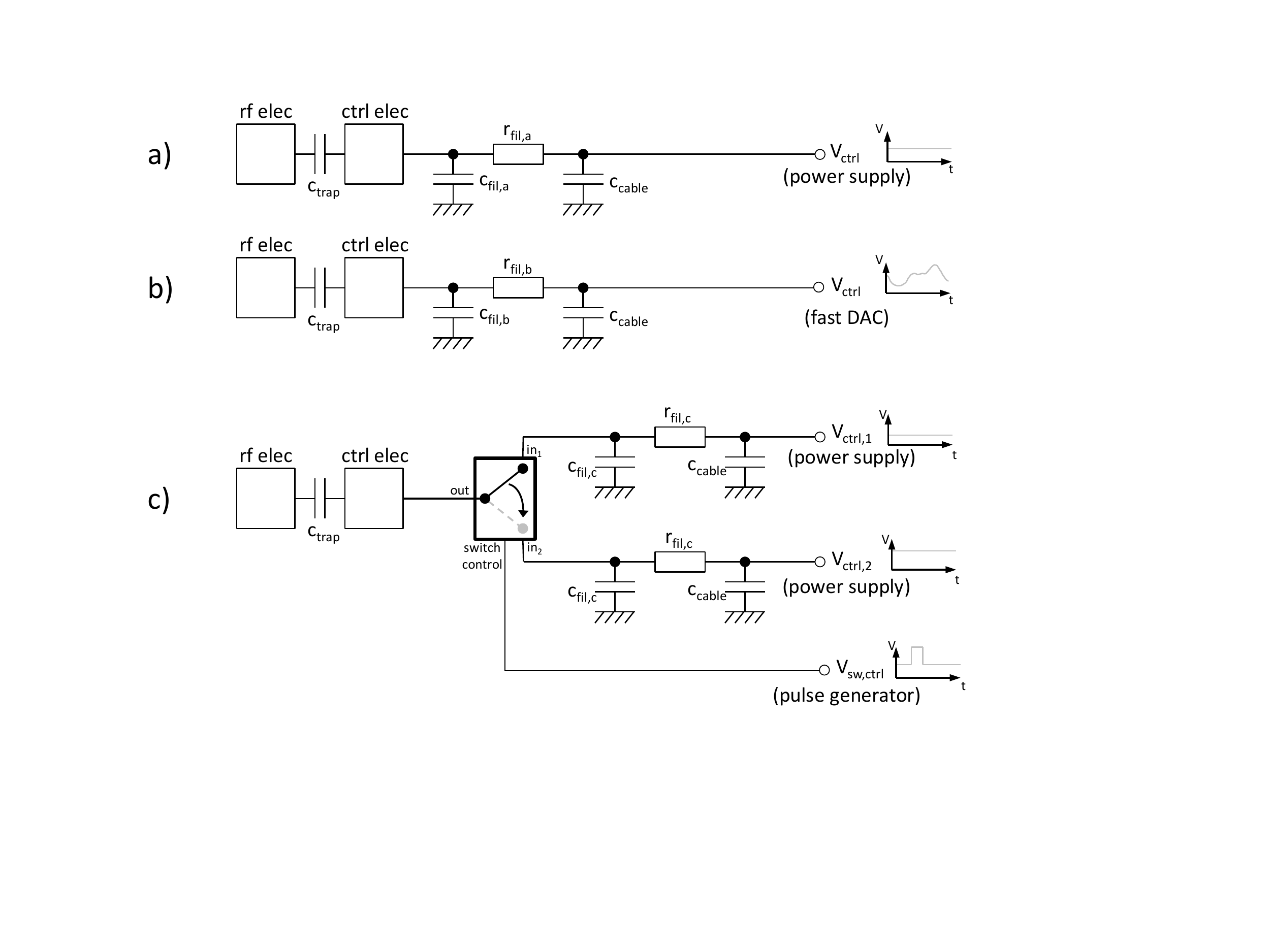}
}
\caption{Electronic schemes for setting the voltages at the control electrodes of a trap. The standard configuration is shown in a) and consists of a voltage supply and a low-frequency filter ($r_\text{fil,a}$ and $c_\text{fil,a}$ can be chosen such that the corner frequency is $f_\text{c}\sim\unit{10}{\hertz}$). The configuration in b) allows for waveforms, typically at frequencies close to but lower than the secular frequency. The idea is the same as in the standard configuration, only the voltage supply is replaced by a digital-to-analog converter and the corner frequency of the low-pass filter is set to a higher value ($f_\text{c}\sim\unit{100}{\kilo\hertz}$). The ultra-fast switching scheme described in the text is shown in c). The values of $r_\text{fil,c}$ and $c_\text{fil,c}$ can be chosen such that $f_\text{c}\sim\unit{10}{\hertz}$. In all cases, the capacitors $c_\text{fil}$ have two functions: they form part of the low-pass filters for preventing high-frequency noise from getting into the trap; and they serve as a low-impedance path for the rf drive to ground. $c_\text{trap}$ is the capacitance between the rf and the control electrodes, and $c_\text{cable}$ represents the parasitic capacitance of the typically long cables and feedthrough assemblies from the external electronics to the trap chamber. The small insets next to the supplies indicate the generic time-dependent voltages on the respective lines.}
\label{fig:rffilter}
\end{figure}

\begin{figure}
\centering
\resizebox{0.75\textwidth}{!}{
  \includegraphics{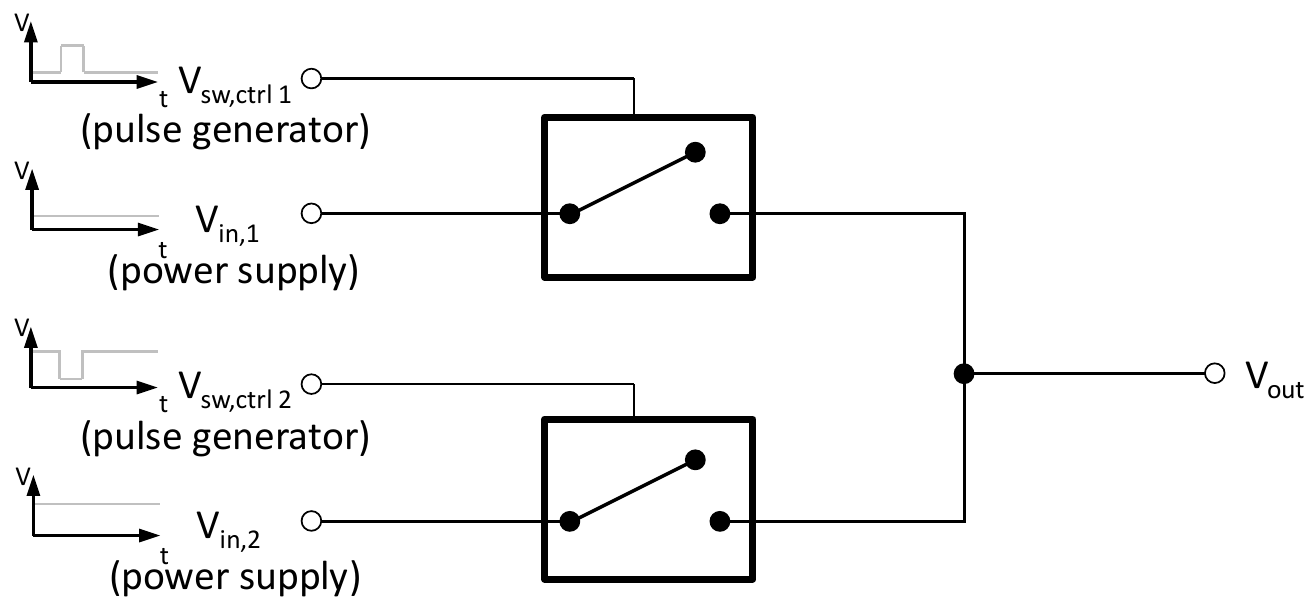}
}
\caption{Simplified SPDT implementation with two SPST switches. This scheme requires proper synchronization of both digital control signals. Further electronic components required for the experimental implementation of the SPDT have been omitted.}
\label{fig:spdt}
\end{figure}

\begin{figure}
\centering
\resizebox{0.75\textwidth}{!}{
  \includegraphics{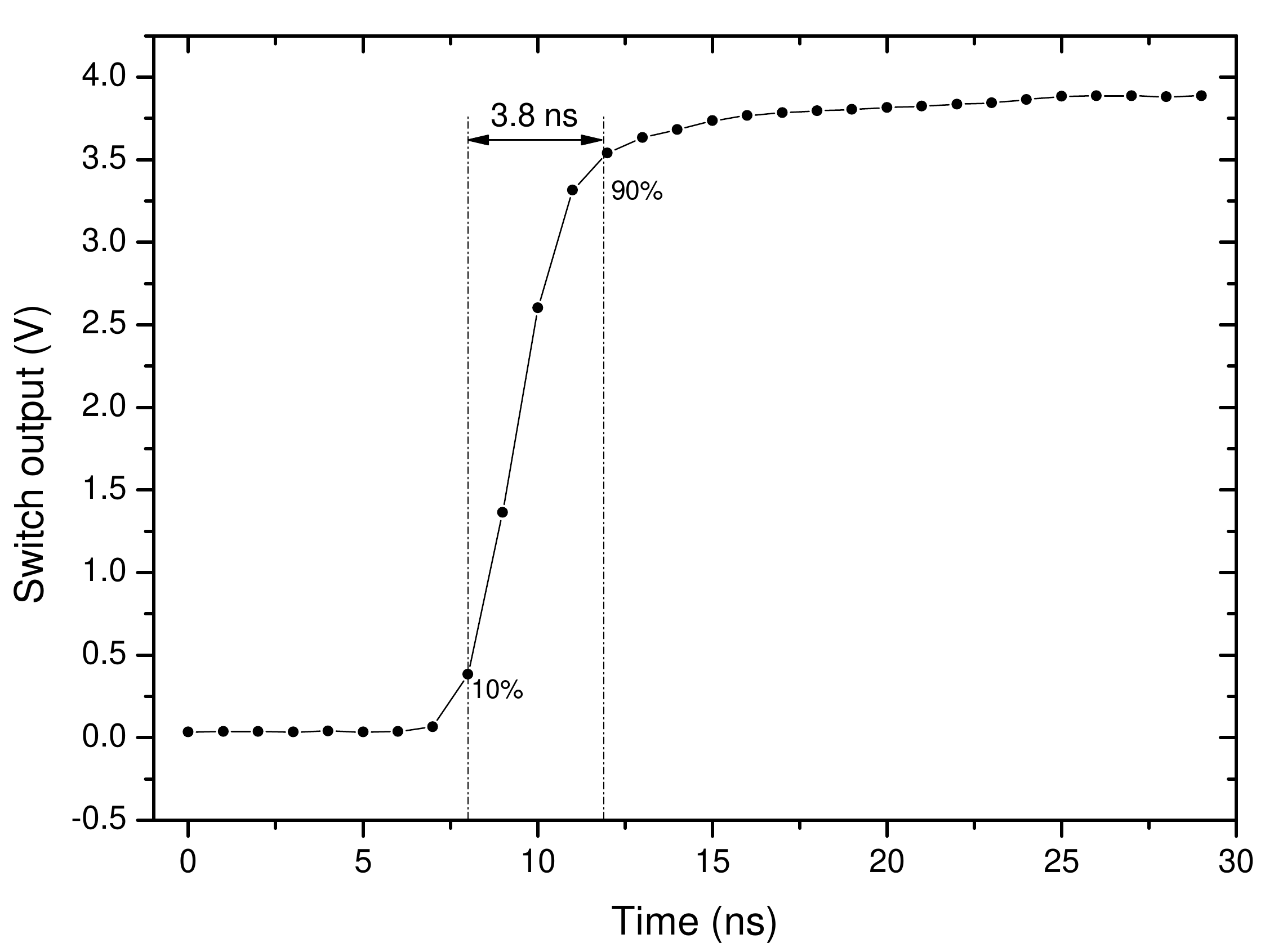}
}
\caption{Switching behavior at \unit{4}{\kelvin} of an SPDT switching circuit based on CMOS SPST switches (74HC4066M from Texas Instruments) following the scheme shown in figure \ref{fig:spdt}. The input voltages are $V_{\text{in},1}=\unit{0}{\volt}$ and $V_{\text{in},2}=\unit{3.8}{V}$. The 10\%-90\% rise time of the switched output is below $\unit{4.0}{\nano\second}$. The upper part of the curve is the expected response due to the load of the detection electronics (the coaxial cables and the oscilloscope introduce a capacitive load of $\sim\unit{50}{\pico\farad}$), which would not be present in the wiring scheme of a trap electrode.}
\label{fig:pulse}
\end{figure}

The goal of our work is to achieve diabatic ultra-fast (for our purposes, this is defined as being faster than $\unit{100}{\mega\hertz}$) control of the trap potential of one or more trapped ions. Additionally, we do not want to relax the requirements on voltage noise on the electrodes, which can lead to ion heating, setting a limit to the coherence of the quantum states of motion and shortening the lifetime of the ion in the trap. The rate of excitation of a single ion from the ground state to the first excited state (which characterizes the rate of heating in a trap) is given by \cite{00Turchette}
\be\label{eq:heatrate}
\Gamma_{0\rightarrow 1} = \frac{q^2}{4 m \omega \hbar} S_E(\omega)\ ,
\ee
where $q$ is the ion charge, $m$ its mass and $\omega$ is the secular frequency in angular units. $S_E(\omega)=2\int_{-\infty}^{\infty} \text{d}\tau \left\langle E(t) E(t + \tau)\right\rangle e^{i \omega \tau}$ is the spectral density of electric field fluctuations in the trap, where $t$ is time and $\left\langle \cdots \right\rangle$ indicates the time-average of the argument. In order to prevent relevant noise spectral components (hundreds of kHz to tens of MHz) from reaching a trapped ion, it is common to place low-pass RC filters close to the ion trap (see figure \ref{fig:rffilter}). These filters attenuate any contribution to $S_E(\omega)$ which comes from outside the trap. However, they also impede the ability to perform fast control of the trapping potentials from outside the vacuum system, since they suppress high frequency components in the analog signals and therefore distort them.

One solution could be to replace the low-pass filters by band-stop filters which allow for both slow and very fast signals. However, parasitic capacitances of the cables to the electrodes (including vacuum feedthroughs) also filter the higher-frequency components and inhibit ultra-fast control. Another possibility is to adjust the waveforms to the filters, such that the voltage control sources deliver high amplitudes of the fast transients and a fraction of them reaches the electrode \cite{12BowlerPrivComm}. In this case, the limit is set by the compromise between the filter strength and the maximum amplitude, as well as the slew rate of the voltage source.

As an alternative to these approaches, we propose to generate the time-dependent control close to the electrodes themselves. The primary ingredient is a Single-Pole Double-Throw (SPDT) switch, the state of which is controlled by a digital line. The SPDT switch provides a connection from one of two input lines to the output dependent on the logical state of the digital input. The chosen input switches abruptly when the digital input reaches a threshold value, thus the precise details of the digital signal arriving at the switch are not important as long as the time at which it reaches the switching value does not fluctuate. The speed at which the voltage switches is determined exclusively by the characteristics of the switch itself and the capacitive load at the output.

Figure \ref{fig:rffilter} c) shows the electronic scheme for wiring up a control electrode to analog inputs in our proposal. The switch is placed between the low-pass filter and the electrode, otherwise the filter would also attenuate the fast components of the switched signal.

A possible implementation of an SPDT switch based on more common Single-Pole Single-Throw (SPST) switches is shown in figure \ref{fig:spdt}. An illustration of the analog output of a candidate SPDT switching circuit based on CMOS SPST switches (74HC4066M chip from Texas Instruments) operating at \unit{4}{\kelvin} is shown in figure \ref{fig:pulse}, taken from tests of a range of technologies which will be published elsewhere \cite{12Alonso}. The digital control signals are not shown, but the rise/fall times are much longer than that of the switched output ($\sim\unit{50}{\nano\second}$). Despite a slower digital signal, the output switches in $\sim\unit{4}{\nano\second}$. Also visible in the signal is a slow component which changes over a time of \unit{15}{\nano\second}. This is due to our measurement setup, including the capacitance of a \unit{1}{\meter} co-axial cable which was used to connect the output of the switch to the high-impedance oscilloscope and other electronics, introducing an extra capacitive load ($\sim\unit{50}{\pico\farad}$).

As a result of our proposed wiring scheme, the route from the electrode to rf ground now includes the internal impedance of the switch on the on state. It is important that this impedance is low to pull the electrode to rf ground. The on-resistance of the 74HC4066M is specified to be $\sim\unit{15}{\ohm}$ at \unit{300}{\kelvin} \cite{74HC4066M}, and is probably smaller at \unit{4}{\kelvin}, for which we found the switching time to be 3 times faster than at room temperature. As a reference, the typical reactance of the capacitors in the RC-filters ranges from \unit{0.5-10}{\ohm} for trapping-drive frequencies of \unit{20-200}{\mega\hertz} and capacitors close to \unit{1}{\nano\farad} (above $\sim\unit{1}{\nano\farad}$ parasitic inductive effects are relevant). These values have been used in previous experiments \cite{filtercs}.

Another relevant aspect is to design the electronics such that the coupling between the digital and analog sides of electronics is minimized and noise on the digital lines does not reach the electrode (cross talk).

A scheme including multiple-throw switches or a series of cascaded SPDT switches (as opposed to the single SPDT switch depicted in figure \ref{fig:rffilter} c), would allow for a finer control of the electrode voltages.

\section{Experimental implementation}
\label{trap}

\begin{figure}
\centering
\resizebox{0.60\textwidth}{!}{  \includegraphics{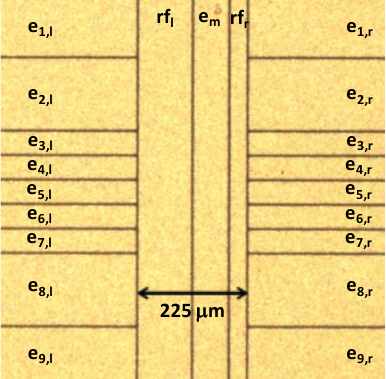}}
\caption{Surface-electrode trap suitable for ultra-fast quantum-control experiments. The rf-electrodes are labeled \textit{rf} and the control electrodes \textit{e}. The main trapping zone is the area sandwiched between the five pairs of narrower (\unit{50}{\micro\meter}) control electrodes, denoted as $e_{2-8}$.}
\label{fig:trap}
\end{figure}

As an illustration of the considerations which must be taken into account when working in this new regime, we next consider the application of our scheme to two tasks in quantum control: diabatic transport of ions and the generation of squeezed states of the motional degrees of freedom of ions. In order to provide an experimental setting, we introduce the fixed geometry of a surface-electrode ion-trap \cite{05Chiaverini,09Amini,09Antohi} which we have recently fabricated. This trap design provides a setting which gives an idea of the size of relevant experimental parameters. However, the methods and routines which we give are general and can in principle be carried out in any other ion-trap setup, so long as there are a sufficient number of independently controllable electrodes. The central region of the trap is shown in figure \ref{fig:trap}. The rf-electrodes and the control electrode $e_\text{m}$ run parallel to the trap axis. By applying an rf voltage with an amplitude of \unit{100}{\volt} at a frequency of \unit{100}{\mega\hertz}, this structure yields a three-dimensional confining pseudo-potential for positively charged particles at a trap-electrode distance of $\sim\unit{60}{\micro\meter}$. The array of control electrodes allows for flexible creation of potential wells along the axis of the trap.

\section{Fast ion transport}
\label{transport}
\begin{figure}
\centering
\resizebox{0.75\textwidth}{!}{
  \includegraphics{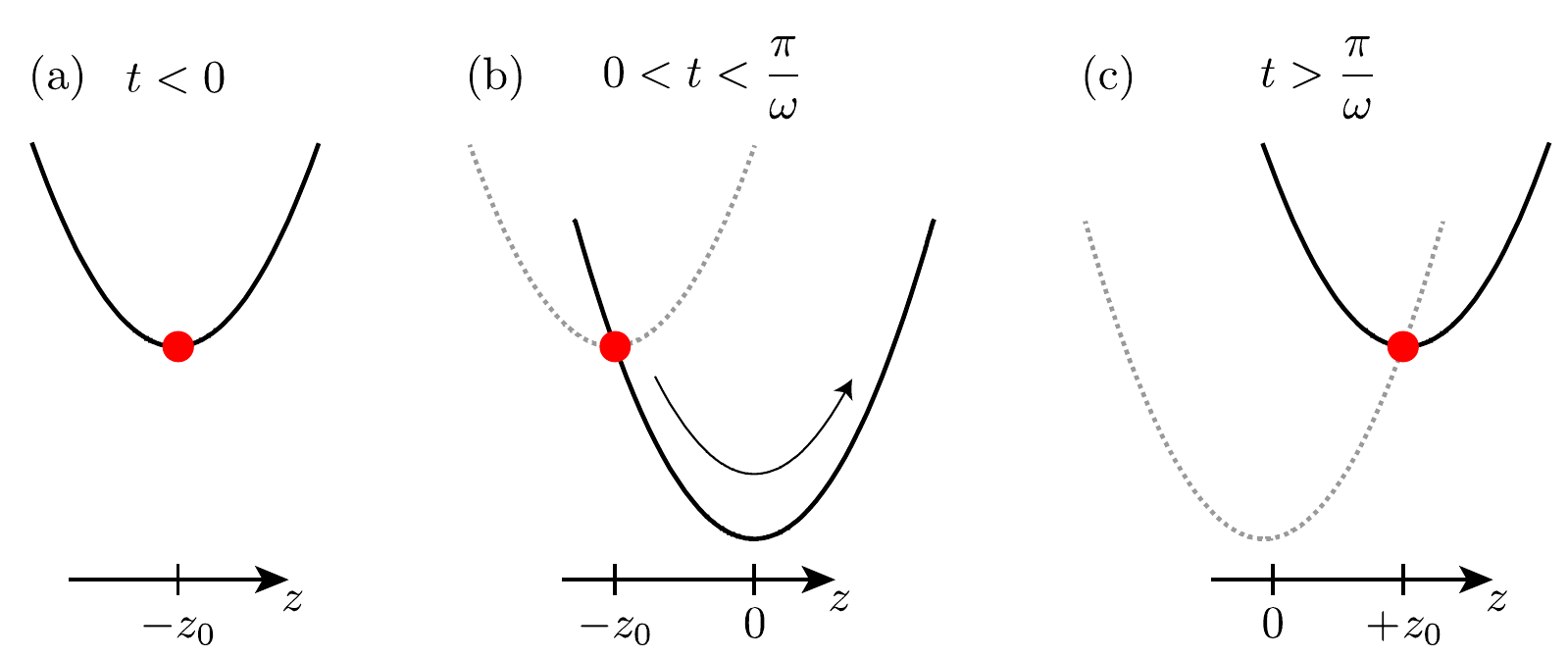}}
\caption{Sketch of the fast ``throw-catch'' transport routine. (a) Starting point: the ion is in the motional ground state of a potential well displaced from the trap center by an offset voltage on an external electrode. (b) Ion transport: the offset voltage is ``instantaneously'' switched off so that the potential is centered in the trap and the ion starts its coherent oscillation. (c) Ion catch: the offset is switched onto a symmetric external electrode and the ion is caught at the instant when it has no kinetic energy, exactly half an oscillation after beginning the transport.}
\label{fig:uftroutine}
\end{figure}

The basic principle of fast transport using sudden switching of potentials is illustrated in figure \ref{fig:uftroutine} for an ion considered a 1-dimensional harmonic oscillator \cite{02Wineland,08Couvert}. The ion starts in the ground state of an initial potential well situated at $z = -z_0$ given by
\be
V_{\rm init}(z) = \frac{1}{2} m \omega^2 (z + z_0)^2.
\ee
At time $t=0$ the potential is suddenly displaced to the transport well, which has the same curvature but is centered at $z = 0$. At this point the ion is in a coherent state of the transport well, with coherent state parameter $\alpha_0=-z_0/a_0$, where $a_0 = \sqrt{\hbar/{2 m \omega}}$ is the r.m.s.~extent of the ground-state wavefunction. Under free evolution, this coherent state will gain and lose momentum, returning to rest periodically at times $t = p \pi/\omega$ where $p$ is an integer. For $p = 1$, the wavepacket is positioned at $z = z_0$. If at this time the potential is suddenly displaced again to a final potential well which is centered at $z = z_0$ and has the same curvature as the transport well, the ion will end up in the ground state of the final potential, having been transported over a distance of $2 z_0$.

Such a ``throw-catch'' transport routine can be carried out in the trap shown in figure \ref{fig:trap}. One possibility is to start with a potential well centered above the pair of electrodes $e_6$ and end above the pair $e_4$. The full transport distance would then be \unit{100}{\micro\meter}. At a typical frequency of \unit{1}{\mega\hertz}, this gives a mean velocity of \unit{200}{\meter/\second} and $\alpha_0\sim4500$ for a \caf ion.

There are a number of experimental challenges behind this conceptually simple transport routine, which we outline below.

\subsection{Micromotion}

Micromotion is the driven motion of trapped ions in rf traps which occurs when the rf component of the electric field is non-zero.

Radial micromotion arises when the dc minimum does not lie on the rf minimum, which is given solely by the geometry of the trap. There are two causes of radial micromotion: stray fields and trapping dc voltages. The former are a priori unpredictable and should be canceled experimentally. In the transport routine, the effect of changing the potential of any single electrode is not only to displace the ion along the $z$-axis, but also in the radial directions. Thus, it will be necessary to switch multiple electrodes simultaneously both in steps (a) and (c) in figure \ref{fig:uftroutine}. The set of voltages required in order to cancel radial micromotion can be determined from simulations \cite{08Wesenberg,09Schmied,10Schmied}. As an example, we give in table \ref{tab:uftparams} the results of such simulations for the trap described in section \ref{trap}.

\setlength{\tabcolsep}{2pt}
\renewcommand{\arraystretch}{1.5}
\begin{table*}\scriptsize
\caption{Example trap voltages for implementing the fast transport routine in the trap shown in figure \ref{fig:trap} with a single $^{40}$Ca$^+$ ion. The transport takes place from $e_6$ to $e_4$ (\unit{100}{\micro\meter}) in half an axial oscillation (\unit{500}{\nano\second}). The largest voltage switch is $<\unit{6}{\volt}$. The energy $E$ of the ion and the relevant potential parameters (defined in section \ref{anharmonic}) are also given. Some electrodes require three different voltages, for which two SPDT switches can be used. Note that the numbers are particular to a specific trap geometry and conditions, but may be used as a reference for different scenarios. The labeling of the electrodes is given in figure \ref{fig:trap}.}
\label{tab:uftparams}       
\begin{center}
\begin{tabular}{|c|c|c|c|c|c|c|c|c|c|c|c|c|c|c|c|c|c|c|}
\hline
 & \multicolumn{13}{|c|}{Control voltages (V)} & \multicolumn{4}{|c|}{Motional parameters} \\
\hline
Step & \multicolumn{2}{|c|}{$e_{1,3,7,9}$} & \multicolumn{2}{|c|}{$e_2$} & \multicolumn{2}{|c|}{$e_4$} & \multicolumn{2}{|c|}{$e_5$} & \multicolumn{2}{|c|}{$e_6$} & \multicolumn{2}{|c|}{$e_8$} & $e_\text{m}$ & $f_\text{z}$ & $L_3$ & $L_4$ & $E$ \\\cline{2-13}
 & l & r & l & r & l & r & l & r & l & r & l & r & & (MHz) & (\textmu m) & (\textmu m) & (eV) \\ \hline\hline
(1) $t\leq0$ & 0.38 & 0.38 & 1.38 & -5.45 & -0.97 & 0.08 & -6.38 & -1.11 & -0.97 & 0.08 & 0.98 & -3.16 & 0.51 & 1.0 & 140 & -200 & $2\cdot10^{-9}$\\
(2) $0<t<\frac{\pi}{\omega}$ & 0.38 & 0.38 & 0.38 & 0.38 & -0.97 & 0.08 & -6.38 & -1.11 & -0.97 & 0.08 & 0.38 & 0.38 & 0 & 1.0 & $10^{10}$ & -120 & $2\cdot10^{-2}$ \\
(3) $t\geq\frac{\pi}{\omega}$ & 0.38 & 0.38 & 0.98 & -3.16 & -0.97 & 0.08 & -6.38 & -1.11 & -0.97 & 0.08 & 1.38 & -5.45 & 0.51 & 1.0 & 140 & -200 & $2\cdot10^{-9}$ \\ \hline
\end{tabular}
\end{center}
\end{table*}

Axial micromotion arises when the axial component of the rf-drive does not vanish, usually due to misalignment, manufacturing imperfections or finite electrode structures. There are two relevant components to axial micromotion: one arising from a residual electric field of homogeneous amplitude along the axis, and the second due to the curvature of the pseudo-potential along the trap axis. For an ion of charge $Q$ and mass $m$, the combined effect leads to a time-dependent motion which can be approximated by \cite{10Blakestad}
\be
x(t)\approx C_{0}\cos(\omega t)+C_{2}\cos[(\Omega+\omega) t]+C_{-2}\cos[(\Omega-\omega) t]+D_{2}\cos(\Omega t),
\ee
where $\Omega$ and $\omega$ are the drive and secular angular frequencies, respectively, $C_{0}$ is the (classical) amplitude of the secular oscillation ($\unit{100}{\micro\meter}$ in the transport example above), $C_{\pm2}$ are due to the curvature of the pseudo-potential and $D_2$ is due to the residual oscillating electric field. The size of $C_{\pm2}$ is given by
\be
C_{\pm2}\approx-C_0\frac{q_z}{a_z-(\pm2+\beta_z)^2},
\ee
with
\be
\nonumber \beta & \approx & \sqrt{a_z+q_z^2/2},\\
\nonumber a_z & = & \frac{8Q}{m\Omega^2}c_2,\\
 q_z & = & \frac{4Q}{m\Omega^2}d_2.
\ee
Here, $c_2$ and $d_2$ are the second order coefficients in the dc and rf contributions to the axial potential, respectively. In our trap, $c_2\simeq\unit{8\cdot10^{-6}}{\volt/\meter^2}$ and $d_2\simeq\unit{4}{\volt/\meter^2}$, so $C_{\pm2}\simeq\unit{2.5}{\femto\meter}$. $D_2$ could be larger, but its effect will time-average to zero during ion transport, since the drive frequency is required to be much higher than the axial secular frequency for stable trapping.

In practice, a careful experimental characterization of the potentials created by the electrodes will be required in situ before the proposed methods can be implemented, because any real electrode structure will be finite, have gaps and be subject to fabrication imperfections.

\subsection{Effect of trap anharmonicity}
\label{anharmonic}

For macroscopic transport distances (\unit{100}{\micro\meter} in the example considered), the anharmonicities experienced by the ion during its coherent oscillation might not be negligible. The trapping potential can be expanded as
\be\label{eq:potential}
V(z)\approx2\pi^2mf_\text{z}^2z^2\left(1+\frac{z}{L_3}+\text{sgn}(L_4)\frac{z^2}{L_4^2}\right),
\ee
where $f_\text{z}$ is the oscillation frequency of the ion (determined from the curvature of the potential), and $L_3$ and $L_4$ are, respectively, the length scales at which the effects of the cubic and quartic terms of the potential are of the same size as the quadratic term. The principal anharmonicity relevant in the dynamics described above is a quartic term which causes a variation of the curvature of the potential between $z = \pm z_0$ and $z = 0$ (the odd term in the expansion in equation \ref{eq:potential} is negligible due to symmetry, see table \ref{tab:uftparams}). The effect of $L_4$ on the frequency can be evaluated by taking the second spatial derivative of the potential in equation \ref{eq:potential}. Simple calculations lead to
\be\label{eq:omegaq}
\omega^2(z)=\omega^2(0)\left(1+6\frac{\text{sgn}(L_4)}{L_4^2}z^2\right).
\ee

The consequences of equation \ref{eq:omegaq} were simulated numerically using a Suzuki-Trotter expansion \cite{93Suzuki} of the full Hamiltonian (split-operator method). Due to the size of the coherent state under consideration, the simulation required a time resolution of \unit{360}{\pico\second} (1400 time steps) and a position resolution of $\sim\unit{8}{\pico\meter}$ to describe the wavepacket (r.m.s.~extent of $\sim\unit{10}{\nano\meter}$).

\begin{figure}
\centering
\resizebox{1\textwidth}{!}{
  \includegraphics{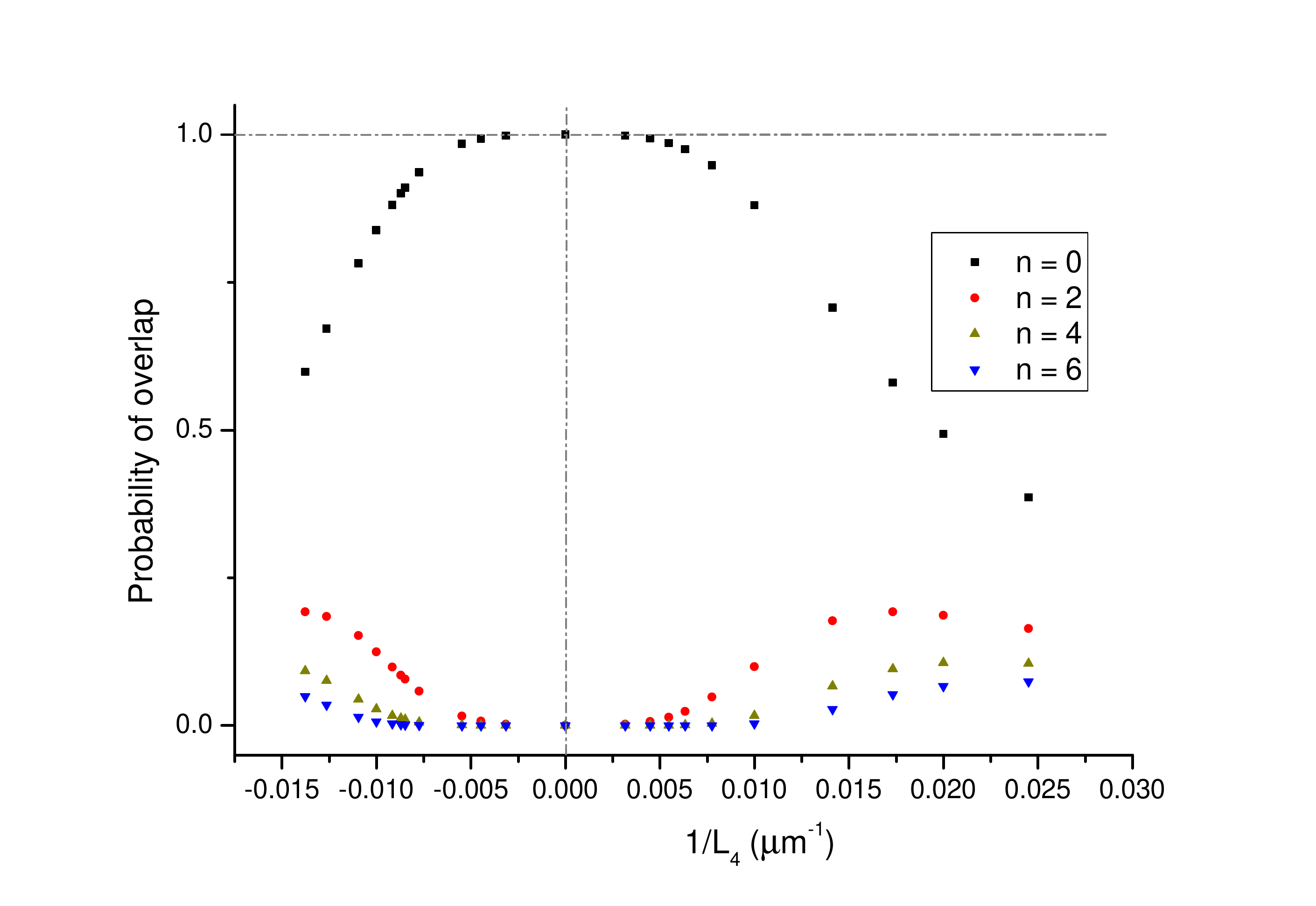}}
\caption{Probability of overlap of the final state with the Fock states $n=0,2,4,6$ of the final potential as a function of the strength of the quartic anharmonic term during transport. The data points have been simulated for a \caf ion at an axial frequency of \unit{1}{\mega\hertz} which undergoes a transport of \unit{100}{\micro\meter} ($z_0=\unit{50}{\micro\meter}$). For $1/L_4<-\unit{0.014}{\micro\meter^{-1}}$, the potential becomes anti-confining at the edges of the transport region. The results of the simulations are independent of the axial frequency to within the numerical uncertainties.}
\label{fig:quartic}
\end{figure}

Figure \ref{fig:quartic} shows the effect of the quartic anharmonicity on the probability of overlap of the final state with the lowest Fock states in the final well. All three potential wells were taken to have the same curvature at their respective centers, leading to an axial frequency of \unit{1}{\mega\hertz}. For negative values of $L_4$, the curvature of the transport potential becomes negative at $|z_1|\geq |L_4|/\sqrt{6}$ and the potential becomes anti-confining at $|z_2|\geq |L_4|/\sqrt{2}$. In general, the final state is no longer a minimum uncertainty state, since the variance in momentum increases. Using the parameters listed in table \ref{tab:uftparams}, the probability of overlap of the state in the catch potential with its ground state is 90\%, and with the $n=2$ state 10\% at $t=\unit{500}{\nano\second}$.

One possibility to minimize this effect is to design the transport well such that the quartic term in the potential is reduced, either using geometry or suitably chosen electrode potentials \cite{06Home}. For the trap geometry described above, we find a combination of voltages which cancels out higher order contributions, however this requires switching over a voltage range which exceeds the capabilities of any device which we have tested. This forms an extra consideration for any chosen implementation.

\begin{figure}
\centering
\resizebox{0.75\textwidth}{!}{
  \includegraphics{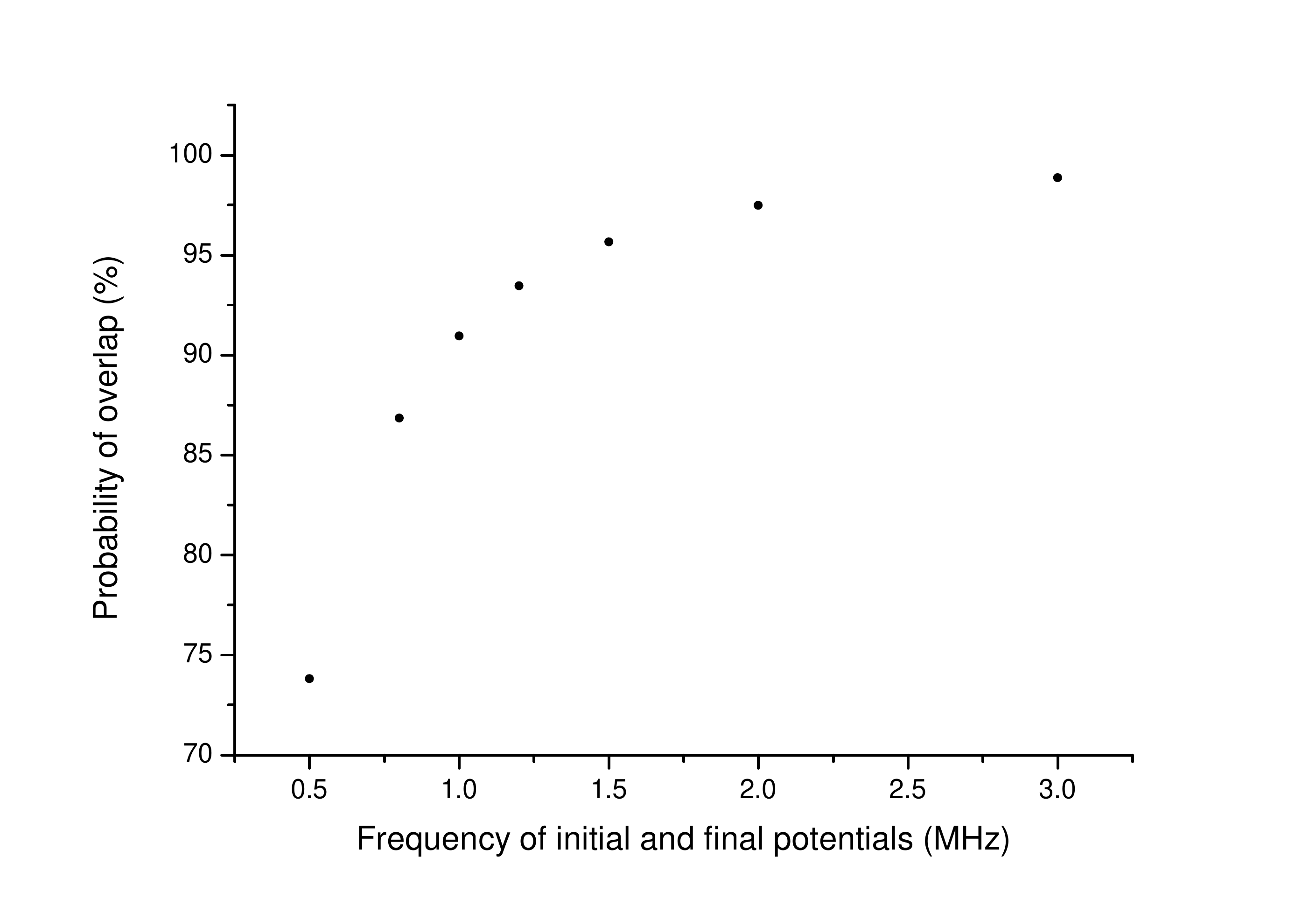}}
\caption{Probability of overlap with the ground state of the final potential as a function of the frequency of the initial and final potentials. The transport potential is fixed so that $\omega(0)=2\pi\cdot(\unit{1}{\mega\hertz})$ and $L_4=\unit{-120}{\micro\meter}$. The total transport distance is \unit{100}{\micro\meter} ($z_0=\unit{50}{\micro\meter}$). The highest frequency simulated was \unit{3.0}{\mega\hertz} due to computational limitations.}
\label{fig:transsq}
\end{figure}

We also investigated having a larger curvature in the initial and final wells than in the transport well. In that case, the ground state of the initial well projects onto a squeezed (rather than coherent) state of the transport well. The probability of overlap with the ground state of the final well is shown in figure \ref{fig:transsq} as a function of the frequency of the initial and final potentials, given a fixed transport potential. The highest frequency simulated was \unit{3.0}{\mega\hertz} due to computational limitations. Until that value, the probability of overlap increases because the effect of the anharmonicity of the transport potential is smaller across the squeezed wave function. It might come as a surprise that the curve seems to asymptotically approach a probability of overlap of 100\%. Indeed, we expect that it will reach maximum. The reason is that, close to the trap center, the wavefunction of the squeezed state broadens. Above a certain frequency, the effect of the anharmonicity on the broadened wavefunction will be large enough to reduce the probability of overlap with the final potential well.

\subsection{Switch timing}

Since the timing of the switch to the final potential well is a critical part of the transport protocol, it is worth examining the tolerance of the scheme to timing imprecision. We can make an estimate by considering that the ion in the transport potential is in a coherent state characterized by the eigenvalue
\begin{equation}
\alpha_0=-\frac{z_0}{a_0}
\end{equation}
of the harmonic oscillator annihilation operator, where $z_0$ is the initial displacement from the midpoint and $a_0=\sqrt{\hbar/(2 m\omega)}$ as before. If the catch potential is switched on at $t_\text{f}=\pi/\omega+\delta t$ (rather than at $t=\pi/\omega$), the ion was in the coherent state
\be
\ket{\alpha_\text{f}}=\ket{-\alpha_\text{0}e^{-i\omega\delta t}}
\ee
of the transport potential. The overlap with the ground state of the final well is then given by
\be
\bra{0}\hat{D}^\dagger(-\alpha_0)\ket{\alpha_\text{f}}=\braket{0}{-\alpha_0\left(-1+e^{-i\omega\delta t}\right)}\simeq\text{exp}\left\{-\frac{1}{2}\left(\alpha_0\omega\delta t\right)^2\right\}.
\ee
where we have approximated $\delta t\ll1/\omega$.

The probability of overlap is $P_\text{O}=\left|\bra{0}\hat{D}^\dagger(-\alpha_0)\ket{\alpha_\text{f}}\right|^2$, so the constraint on $\delta t$ in order to achieve an overlap of $P_\text{O}$ is
\be
\delta t=\frac{\sqrt{-\ln P_\text{O}}}{|\alpha_0|\omega}.
\ee

As a reference, for a $^{40}$Ca$^+$ ion at $\omega=2\pi\cdot(\unit{1}{\mega\hertz})$ and $z_0=\unit{50}{\micro\meter}$, the r.m.s.~extent of the initial ground-state wavefunction is $a_0\simeq\unit{11}{\nano\meter}$, the initial coherent state is $|\alpha_0|\simeq 4450$ and the timing resolution required is $\delta t\leq\unit{12}{\pico\second}$ for $P_\text{O}\geq0.9$. This poses a demanding constraint on the trapping potential, which should be stable to $\sim\unit{10}{\pico\second}/\unit{1}{\micro\second}=10^{-5}$ throughout a measurement to avoid recalibration of the digital delays within an experiment. All active electronics as well as the delays due to impedances in the connections to the trap and the trap itself must also be stable at this level so that they can be calibrated out. Timing resolutions on the order of tens of picoseconds are achievable with currently available electronics  \cite{picosec,12Alonso}.

\subsection{Finite switching time}
\label{finite}

\begin{figure}
\centering
\resizebox{0.75\textwidth}{!}{
  \includegraphics{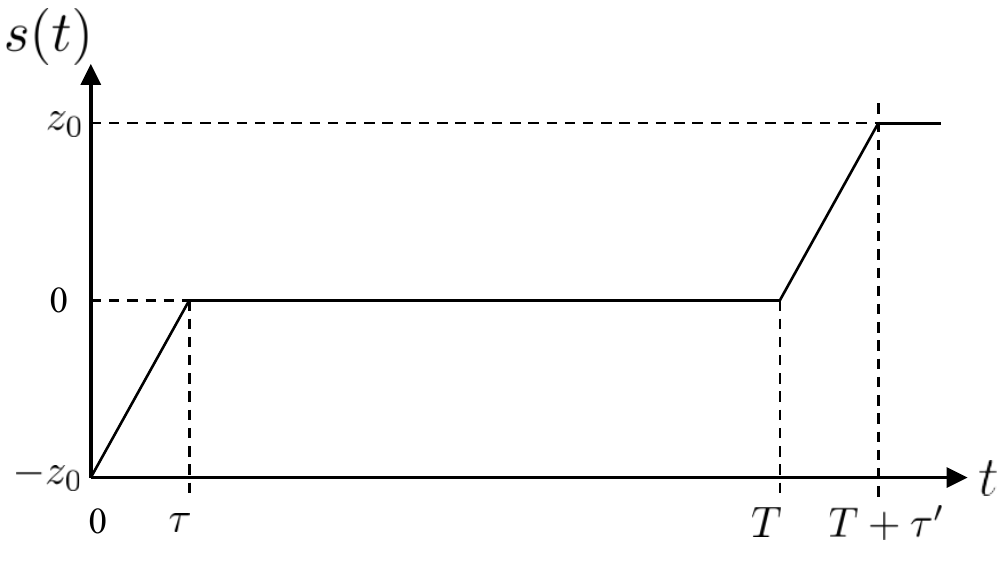}}
\caption{Position of the potential minimum as a function of time during transport. The transitions from 0 to $\tau$ and from $T$ to $T+\tau'$ will be given by the experimental setup, and in general not be linear.}
\label{fig:finiteswitch}
\end{figure}

Until now, we have assumed that the voltage switching is instantaneous. In our proposed transport scheme, the ion is transported by two subsequent displacements of a constant-curvature harmonic potential, which we can describe by $\frac{1}{2}m\omega^2(z-s(t))^2$, where $s(t)$ is the position of the potential minimum. Finite switching times of electrode potentials will mean that the potential minimum takes a certain time $\tau$ to move from the initial position $s(0)=-z_0$ to the transport well centered at $s(\tau)=0$. The same applies to the catch procedure, which will start at $s(T)=0$ and will end at $s(T+\tau')=z_0$ (see figure \ref{fig:finiteswitch}). Note that $\tau$ and $\tau'$ are related to the switching time of the electrode potentials, but may slightly differ from it if more than one electrode is switched.

In \cite{11Lau}, it is shown that if the ion is initially in the ground state of motion on the starting potential $s(0)$, it will evolve into a coherent state with amplitude
\be
\alpha(t)=-\sqrt{\frac{m\omega}{2\hbar}}e^{-i\omega t}\int_0^t\text{d} t'\dot{s}(t')e^{i\omega t'}
\ee
with respect to the frame centered at $s(t)$. For our case, this means that the ion will finish in a coherent state with amplitude
\be\label{eq:finswitch1}
\nonumber \alpha(T+\tau')& = &-\sqrt{\frac{m\omega}{2\hbar}}e^{-i\omega(T+\tau')}\left[\int_0^\tau\text{d}t'\dot{s}(t')e^{i\omega t'}\right.\\
& & +\left.\int_0^{\tau'}\text{d}t'\dot{s}(t'+T)e^{i\omega(t'+T)}\right]
\ee
with respect to the frame centered at $s(T+\tau')$.

From equation \ref{eq:finswitch1} it can be seen that the effects of finite switching times can be canceled out, i.e. it is possible to catch the ion in the ground state of a final potential, at the expense of having $T$ and $s(T+\tau')$ as variable parameters. However, it is in general not possible to make the transport symmetric, i.e. fix $s(T+\tau')=-s(0)$, and at the same time catch the ion in the ground state by controlling only $T$.

\begin{figure}
\centering
\resizebox{0.75\textwidth}{!}{
  \includegraphics{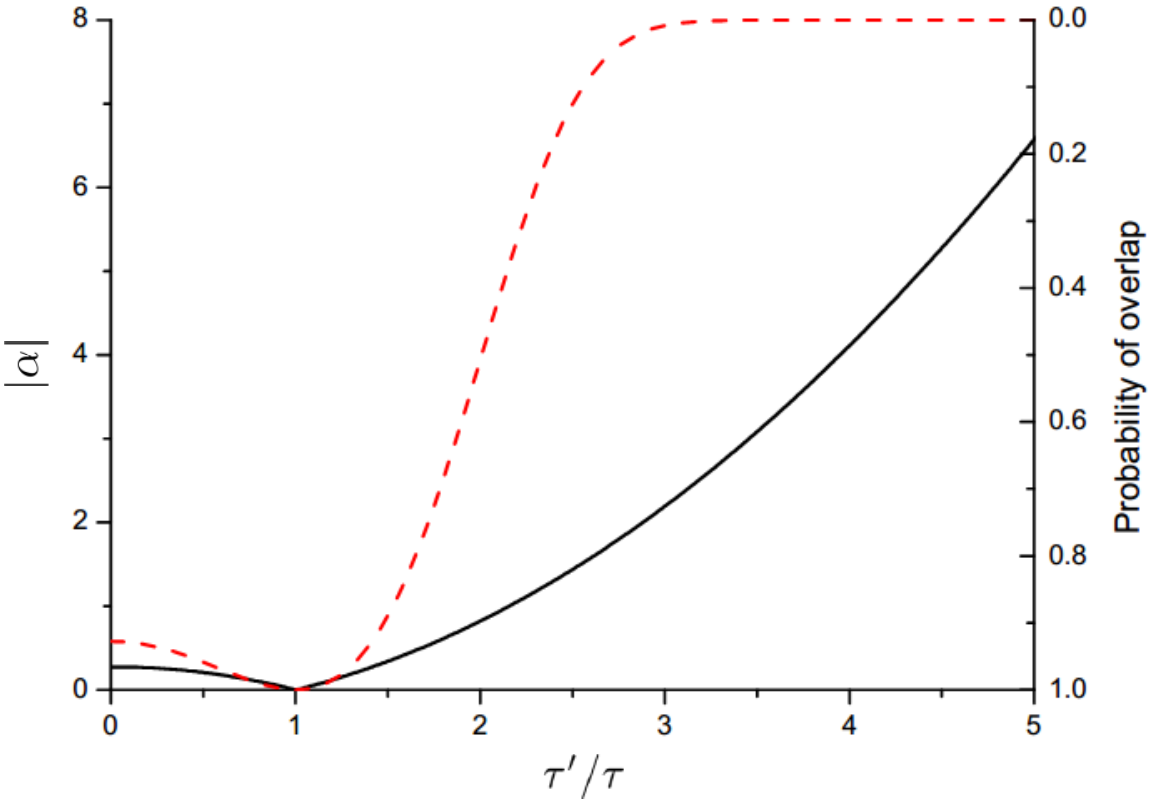}}
\caption{Black solid line (left axis): minimum size of the final coherent state $|\alpha|$ as a function of $\tau'$. Red dashed line (right axis): probability of overlap of the final coherent state with the ground state of the final potential. Data calculated for a \caf ion at $\omega=2\pi\cdot(\unit{1}{\mega\hertz})$, $z_0=\unit{50}{\micro\meter}$ and $\tau=\unit{5}{\nano\second}$. The displacements of the potential minima are assumed to be linear.}
\label{fig:finswtchalpha}
\end{figure}

In order to estimate the size of the effects of finite switching times, we performed analytical calculations (see \ref{linsin}) for the case of linear and sinusoidal transitions. For realistic experimental parameters their effect is very similar. As an illustration, the size of the final coherent state for the linear case (equation \ref{eq:finswitch2}) is shown as a function of $\tau'$ in figure \ref{fig:finswtchalpha} for a \caf ion at $\omega=2\pi\cdot(\unit{1}{\mega\hertz})$, $z_0=\unit{50}{\micro\meter}$ and $\tau=\unit{5}{\nano\second}$. In such a setup, to catch the ion with a probability of overlap above 90\% with the ground-state wavefunction, it suffices to have $\tau'<1.5\tau$.

For an arbitrarily shaped transition, if $s(T+\tau')=z_0$, $T=\pi/\omega$ and $\tau'=\tau$, then equation \ref{eq:finswitch1} becomes
\be
\alpha(T+\tau)=\sqrt{\frac{m\omega}{2\hbar}}e^{-i\omega\tau}\int_0^\tau\text{d} t'\left[\dot{s}(t')-\dot{s}(t'+T)\right]e^{i\omega t'}.
\ee
Hence, if the transitions are equivalent, i.e. $\dot{s}(t')=\dot{s}(t'+T)$, we can recover the ion in the motional ground state after transport, making the process first-order insensitive to finite switching times.

\subsection{Extended transport throughout an array}
\label{scaling}

\begin{figure}
\centering
\resizebox{1\textwidth}{!}{
  \includegraphics{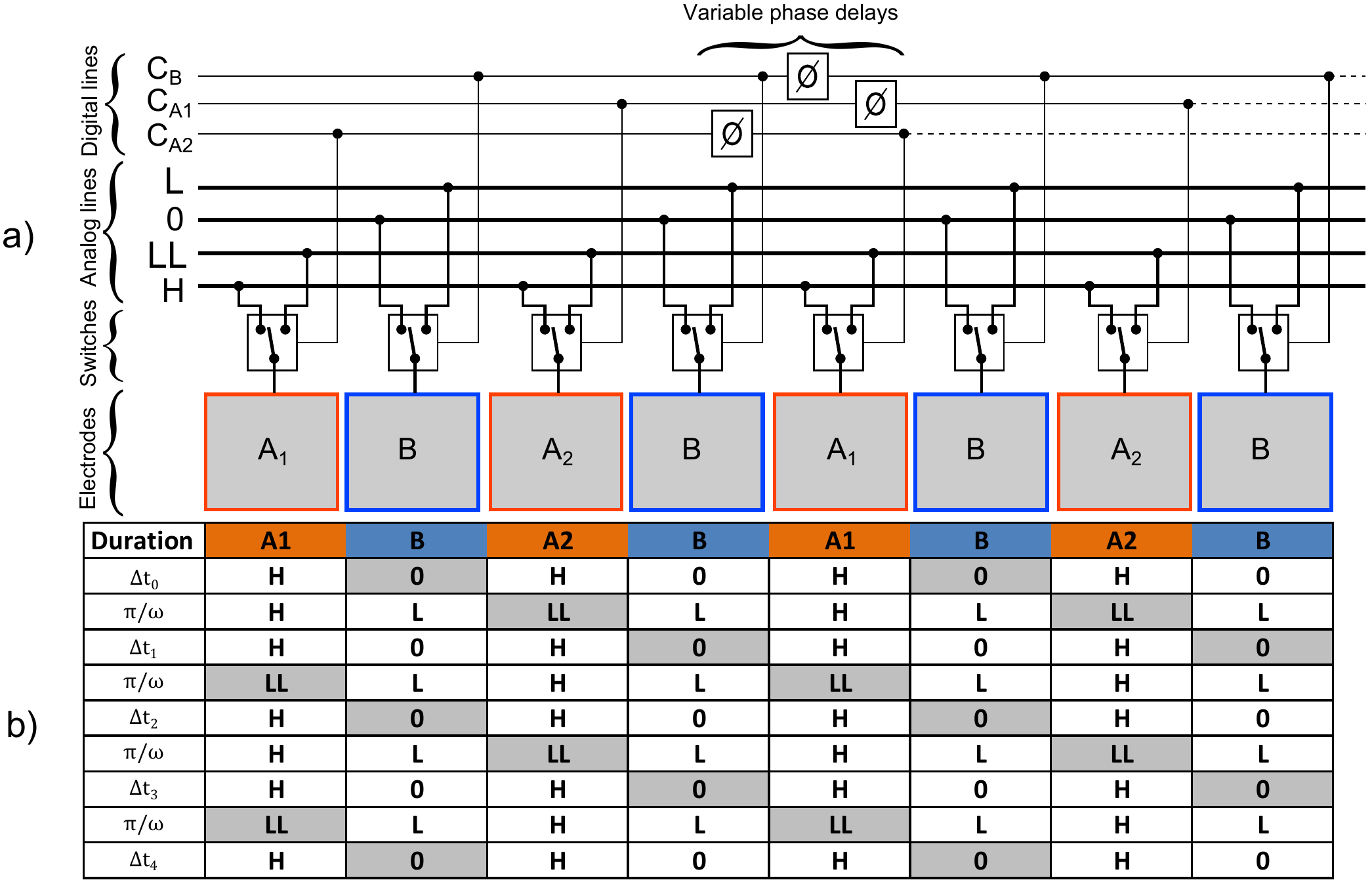}}
\caption{a) Connections required to switch multiple trapping regions, in an ideal case. One row of electrodes and connections to the bus are shown. The ions are separated by three electrodes in this scheme. Electrodes A are switched between voltages H and LL, and electrodes B between 0 and L, with H$>$0$>$L$>$LL. Control lines C$_\textrm{A}$ and C$_\textrm{B}$ switch the voltages applied to the pairs of electrodes A and B, respectively. b) Switching steps for the transport operations. The ions' positions (centered above an electrode) are indicated by a shaded cell. It allows an ion to be moved two electrodes to the right with a time period $\pi/\omega$. $\Delta t_i$ account for the time delays during which the ions are in the ground state of the potential wells centered above the B electrodes. $\Delta t_i=0$ would be desirable for fast transport, while finite values would allow for operations to be carried out on the ions.}
\label{fig:fabric}
\end{figure}

In a large-scale trapped-ion quantum information processor based on the architecture described by Kielpinski \emph{et al.} \cite{02Kielpinski}, it is necessary to move ions between many remote trapping regions. Therefore it is worth considering how the scheme outlined above could be extended and implemented in a repeated fashion. Here, experimental simplicity means that it is desirable to minimize the number of connections which are required. In figure \ref{fig:fabric} we give a simplified outline of a switch fabric which could be used to wire up electrodes in order to implement transport between a number of adjacent zones of a multi-zone trap. Though this does not include all electrodes required to control transport over any single zone, it illustrates that increasing the number of zones does not imply a direct increase in the number of digital and analog control lines, since repeated transport in this scheme is realized by a total of four analog voltages controlled by three digital lines.

In a real experiment, it is likely that fabrication imperfections and stray environmental electric fields would require additional electrodes which could be used to tune the trap frequency of the ion. In addition, switch timing could be controlled by adding phase-shifting elements such as varactors, which have previously been proven to work in cryogenic ion trap settings \cite{12Schabinger}.

With a chain of ions of equal mass, it should be possible to shunt the whole string along by one trap zone by switching the electrodes in the same manner as for a single ion. This is due to the fact that, as long as the curvature of the potential seen by the ions during the transport process remains the same, no motional modes will be affected other than the center-of-mass mode, which is excited exactly as a single ion would be \cite{98James}. In addition to transport, in a large-scale processor it is likely that the ability to separate and deterministically re-order ion strings will be necessary. It is conceivable that methods similar to those we have described above could be used for these tasks. Separation and recombination will inevitably involve anharmonicity in the potential (trapping and Coulomb repulsion), which may require methods such as those we describe in section \ref{anharmonic} to keep the ions close to the ground state.

\section{Motional-state squeezing}
\label{squeezing}

An alternative application of the fast switching method is in the generation of non-classical states. Previous work by Serafini and co-workers describes methods for generating continuous-variable entanglement by applying ultra-fast control of the radial degrees of freedom of an ion chain \cite{09Serafini,09Serafini2}. A critical ingredient of the Serafini scheme is the generation of squeezing of one of the ions, with the Coulomb interaction between ions then giving rise to entanglement. As a first step towards such experiments, we consider the lifetime of the states squeezed by applying ultra-fast control of the trapping potential, which is limited by the presence of a quartic term in the potential as well as heating due to fluctuating electric fields.

\begin{figure}
\centering
\resizebox{1\textwidth}{!}{
  \includegraphics{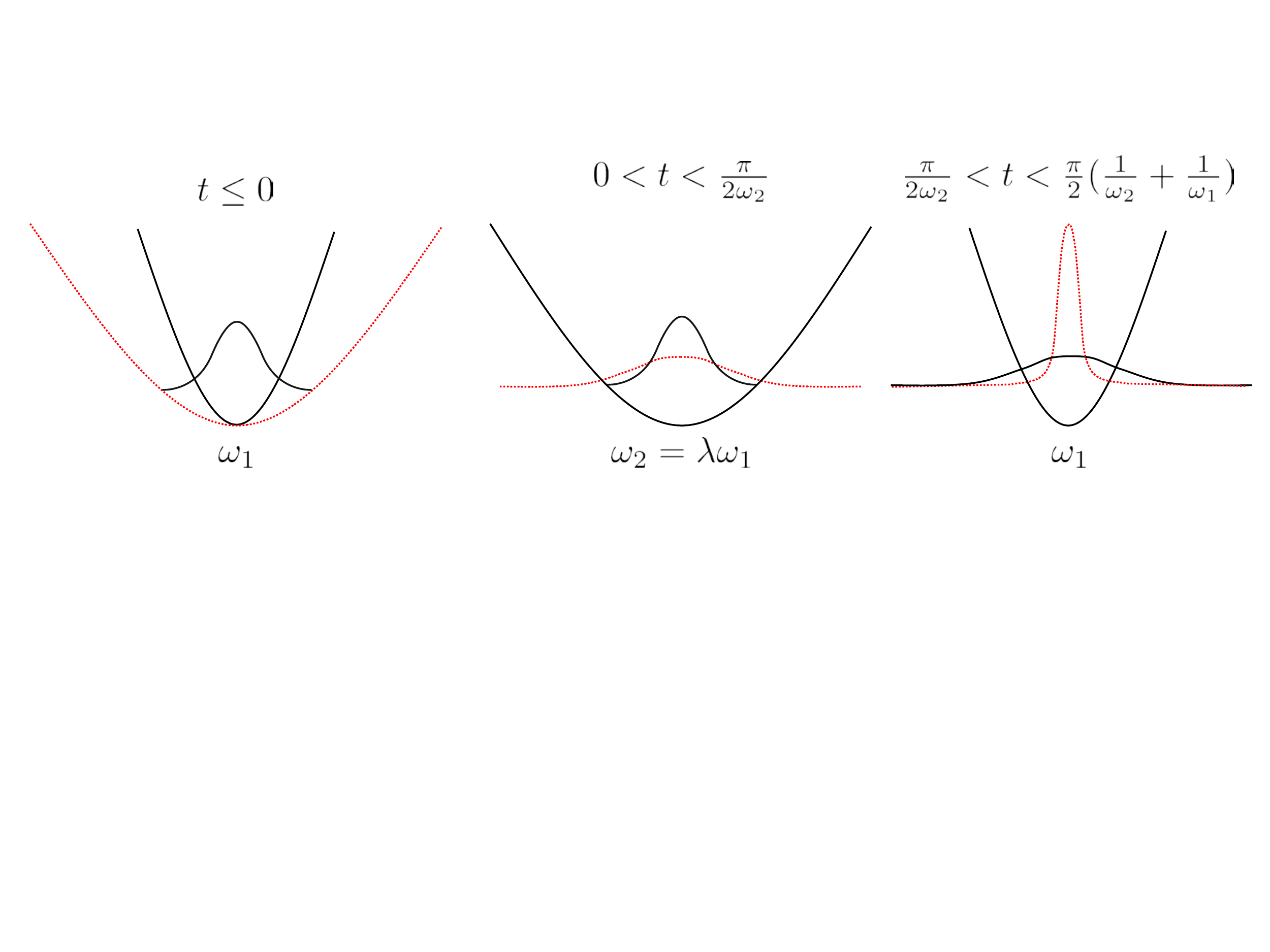}}
\caption{Illustration of the squeezing mechanism based on voltage switching. The ion starts in the motional ground state of the first potential and at $t=0$ its curvature is suddenly switched. The wave function starts breathing periodically with angular frequency $2\omega_2$, accommodating to the new potential. At $t=\pi/(2\omega_2)$ it reaches a maximum width and the potential is switched back, leaving the state squeezed. Black solid lines represent the wave function and the potential at the beginning of the period indicated and red dashed lines at the end of it.}
\label{fig:sqscheme}
\end{figure}

Squeezing refers to a state in which quantum fluctuations in one quadrature are suppressed relative to their value in the ground state \cite{88Henry}. The orthogonal quadrature has increased fluctuations, preserving the Heisenberg uncertainty relation. Mathematically, a squeezed state can be obtained from the vacuum state by applying the operator
\be
\hat{\Sigma}(\xi)=e^{\left(\xi^*(\destroy)^2-\xi(\create)^2\right)/2},
\ee
where the parameter $\xi = r e^{i \Phi}$ is the complex squeezing parameter, denoted by a real magnitude $r$ and phase $\Phi$. The squeezed state is one example of a Gaussian continuous variable state, which can be characterized in terms of the first moments $\text{Tr}[\hat{R_j}\rho]$, and the second moments, embodied in the covariance matrix $\sigma$ \cite{10Cormick,03Eisert}
\be
\sigma_{jk}=\frac{1}{2}\text{Tr}[\{\hat{R_j},\hat{R_k}\}\rho]-\text{Tr}[\hat{R_j}\rho]\text{Tr}[\hat{R_k}\rho].
\ee
Here, $\rho$ is the Gaussian state and $\hat{R}$ is a vector of normalized position and momentum operators. For one ion, $\hat{R}=(\hat{x},\hat{p})^\text{T}$. The amount of squeezing of the state is given by the smallest eigenvalue of the covariance matrix. For details, see e.g. \cite{03Wolf}.

Squeezing of motional states of trapped ions has already been reported using optical fields to create a parametric drive which squeezed the vacuum state \cite{96Meekhof}. Instead, we consider the squeezing procedure proposed in \cite{09Serafini}. It involves an abrupt change of the trapping frequency from $\omega$ to $\lambda\omega$ at time $t = 0$, followed by the opposite transformation at time $t = \pi/(2\lambda \omega)$ (see figure \ref{fig:sqscheme}). For a Gaussian initial state with mean zero, the time evolution can be fully characterized by the transformation of the covariance matrix according to $\sigma_t = S(t) \sigma S(t)^\text{T}$, where \cite{09Serafini}
\be
S(t) = \left(\begin{array}{cc} \cos(\omega \lambda t) & \lambda\sin(\omega \lambda t) \\  -\lambda^{-1} \sin(\omega \lambda t) & \cos(\omega \lambda t) \end{array}\right)
\ee
is a symplectic transformation. For $t = \pi/(2 \lambda \omega)$ we find
\be \label{eq:cmt}
\sigma_t = \left( \begin{array}{cc} \lambda^2 \sigma_{22} & -\sigma_{21} \\ -\sigma_{12} & \sigma_{11}/\lambda^2 \end{array} \right),
\ee
i.e. the variance of one quadrature is squeezed by a factor $\lambda^2$ if $\lambda<1$, while the other increases by the inverse amount.

The major and minor quadrature normalized variances can be shown to be $e^{2r}\hbar/2$ and $e^{-2r}\hbar/2$, respectively \cite{97Barnett}. Therefore, the maximum r.m.s. extent of the squeezed-state wavefunction for a trapped ion is simply $\Delta x=a_0e^{r}=a_0/\lambda$, where the last equality comes from equation \ref{eq:cmt}.

The trap frequency is proportional to the square root of the voltage applied to the central trap electrodes ($e_5$ in figure \ref{fig:trap}), given that the rest of electrodes are grounded. For the trap described above, switching the trapping voltage from \unit{10}{\volt} ($\omega_{1}\simeq2\pi\cdot(\unit{1.2}{\mega\hertz})$) to \unit{1}{\volt} ($\omega_{2}\simeq2\pi\cdot(\unit{400}{\kilo\hertz})$), waiting for a time of $\pi/(2\omega_2)\simeq\unit{625}{\nano\second}$ and switching back to $\omega_{1}$, would give $\lambda^2=10$ vacuum units or \unit{-10}{dB} ($\lambda^2=\unit{0}{dB}$ corresponds to the non-squeezed case). Furthermore, the wavefunction of a squeezed state breathes at $2\omega_{1}$, so the squeezing procedure described above can be repeated with a period $\Delta t=\pi/(2\omega_2)+\pi/(2\omega_1)=(1+1/\lambda)\pi/(2\omega_1)$ to enhance the results.

In the following, we consider the two main experimental limitations to such a procedure: motional heating of the ion in the trap, and anharmonicities in the trapping potential.

\subsection{Motional heating}

An important factor which limits quantum-state engineering of the motional states of trapped ions is heating due to fluctuating electric fields at the position of the ion \cite{11Hite,11Allcock}. In order to estimate the degree to which heating might limit squeezing using the methods discussed in the previous section, we consider the effect of a time-varying electric field $E(t)$ on a squeezed state. In \ref{app} (equation \ref{eq:heatrate2}) we show that the loss of fidelity with the original squeezed state is characterized by a time constant $\tau$, which we can rewrite as
\be
\tau^{-1}=\Gamma_{0\rightarrow 1}\frac{\cosh(2r)}{2}\approx\Gamma_{0\rightarrow 1}\frac{e^{2r}}{4}=\frac{\Gamma_{0\rightarrow 1}}{4\lambda^2}\label{eq:sqheating},
\ee
where $\Gamma_{0\rightarrow 1}$ is given in equation \ref{eq:heatrate} and the approximation is valid for $r\gg1$.

Extrapolating heating-rate measurements in \cite{08Labaziewicz} to the temperature and ion-electrode distance for the trap in figure \ref{fig:trap} yields heating rates down to $\Gamma_{0\rightarrow 1}\sim\unit{10}{quanta/\second}$. For $\lambda^2\sim\unit{-20}{dB}$, which is \unit{10}{dB} stronger than the maximum squeezing obtained with other continuous-quantum-variable systems to date \cite{08Vahlbruch}, equation \ref{eq:sqheating} yields a lifetime of the squeezed state of $\sim\unit{4}{\milli\second}$, much longer than the creation time for this state with our scheme.

\subsection{Effect of anharmonicity}

Anharmonicity in the trapping potential presents another limit to the amount of squeezing which could be achieved. Finite quartic terms in the potential are the most relevant anharmonic contribution for a symmetric trap (see table \ref{tab:uftparams}) and result in an effective spatial frequency distribution $\omega(z)$ as given in equation \ref{eq:omegaq}.

In the phase-space representation given by the Wigner distribution \cite{97Scully}, a squeezed vacuum state takes a two-dimensional Gaussian form centered at the origin \cite{97Breitenbach}. If we let it evolve in time in a harmonic potential, it will rotate around the origin at the trap angular frequency $\omega$. However, in an anharmonic potential the wave function will distort. To quantify the lifetime of the squeezed state in the anharmonic trap, we can calculate the overlap between its wave function and the one it would have become had it evolved in a harmonic potential:
\be\label{eq:overlapsqanhar}
P_\text{O}(t)=\left|\int\Psi^*_\text{anharm}(z,t)\Psi_\text{harm}(z,t)\text{d}z\right|^2.
\ee

We have used the same numerical techniques as in section \ref{anharmonic} to carry out simulations for the lifetimes of squeezed states. For a \caf ion at $\omega=2\pi\cdot(\unit{1}{\mega\hertz})$ with $\lambda^2=\unit{-20}{dB}$ and $L_4=\unit{-120}{\micro\meter}$, the lifetime is $\sim\unit{3}{\milli\second}$. This is comparable to the lifetime expected from the motional heating effects considered in the previous section. However, as discussed in section \ref{anharmonic}, it is possible to improve on the harmonicity of the trapping potential by optimizing the trap geometry and the trapping potentials, whereas reducing motional heating effects is experimentally very challenging. Therefore, we expect that heating effects, not anharmonicities in the potential, will ultimately set the limit to the lifetime of the squeezed motional states in the trap.

\section{Outlook}
\label{outlook}
In the following, we outline a range of further applications which could profit from ultra-fast motional control.

\subsection{Entanglement between radial motional modes}
\label{entanglement}

Linear couplings between two harmonic oscillator modes which are prepared in Gaussian states squeezed along orthogonal quadratures can give rise to entanglement \cite{03Eisert,05Braunstein}. In the case of a string of trapped ions, the motional oscillations of the individual ions are coupled by the Coulomb interaction, for which the lowest order term in the expansion of the interaction potential energy is
\be
\frac{q^2}{4 \pi \epsilon_0 d^3}\chi \xi_1 \xi_2,
\ee
where $\xi_1$ and $\xi_2$ are the excursions of the ion about equilibrium in a given direction, and $\chi = 2$ if the oscillation is aligned with the vector connecting the equilibrium positions of the two ions and $\chi = 1$ otherwise. For two identical ions trapped in the same potential well, the distance between the ions is
\be
d = \left(\frac{q^2}{2 \pi m\epsilon_0 \omega_z^2} \right)^{1/3},
\ee
where $m$ is the mass of a single ion.

Entanglement can be characterized when the two subsystems which are entangled can be accessed locally on timescales which are fast compared to the interaction between these systems. For the case of ions trapped in a single potential well, the interaction energy can be characterized by the splitting of the normal modes, which is typically on the order of a few hundred kHz, and thus the local access to variables (in this case changes in the potential) must happen on timescales short compared to \unit{1}{\micro\second}. Once the voltages are set back to the initial configuration, the squeezing will be transferred from local quadratures to a non-local combination of the coordinates, which will lead to the entanglement.

A similar scheme has been proposed for entangling the modes of nanoelectromechanical oscillators by suddenly switching on an electric interaction between them \cite{04Eisert}.

\subsection{Dynamical studies}
\label{dynamics}

A trapped ion is subject to ``kicks'' due to elastic collisions with background-gas atoms and molecules. In order to understand the mechanisms that lead to Paul-trap instabilities, it might be interesting to map the effect of such kicks as a function of the collisional energy and the resulting impulse direction of the ion.

Collisions could be emulated by pulsing on electric fields. The effect of such collisions can be created in our trap with a routine that starts by cooling an ion to the motional ground state and then adding a sudden offset potential to one or more electrodes, in order to displace the potential minimum and kick the ion in a certain direction. If the ion is left to evolve in the new potential for a time $\pi/(2\omega)$, it will acquire the maximum kinetic energy. At that moment, suddenly removing the offset potential would leave the ion with a high kinetic energy in the original potential well. Another possibility would be to let the ion evolve for a time $\pi/\omega$ in the second well, and then remove the offset to leave the ion with a higher potential energy in the final well.

At room-temperature, the mean thermal energy of the background-gas particles is $E_\text{th}\sim\unit{25}{\milli\electronvolt}$, so the offset voltages required for the routine above are very similar to those required for the ion transport routine (voltages below \unit{10}{\volt} on electrodes $e_{2-8}$ are enough for energy kicks above \unit{20}{\milli\electronvolt}, see table \ref{tab:uftparams}).

\subsection{Interface with solid-state devices}
\label{solidstate}

Fast voltage switching could also be used to help in realizing an interface between a trapped ion and a solid-state quantum device, which could be used to map quantum information from a solid-state qubit to an ion and vice-versa \cite{04Tian}. In order to enhance the coupling, it is desirable to place the ion close to the solid-state device, where anomalous heating could be significant. A fast transport routine similar to that described in section \ref{transport} could be of use to quickly bring the ion close to and away from the device, thus limiting heating effects which occur outside the desired interaction time.

\section{Conclusion}
\label{summary}

We have proposed a method for controlling the potentials in a radio-frequency ion trap on nanosecond timescales. This would allow for ultra-fast control of the potentials seen by the trapped ions, i.e. control at rates much faster than the ions' secular oscillation frequencies. The switching time in our proposal is limited by the capacitive load on the semiconductor-based voltage switches used. We have experimentally measured this to be around \unit{4}{\nano\second} in a 4~Kelvin test apparatus with a load of $\sim\unit{50}{\pico\farad}$.

One possibility opened up by this new approach is to transport an ion within a single oscillation cycle. We have shown that this is experimentally challenging but should be possible for transporting a single \caf ion at an axial angular frequency $\omega=2\pi\cdot(\unit{1}{\mega\hertz})$ over a distance of \unit{100}{\micro\meter} in \unit{500}{\nano\second}.

Another application is the generation of motional squeezed states, where squeezing the ground wavefunction of a \caf ion at $\omega=2\pi\cdot(\unit{1}{\mega\hertz})$ by a factor $\lambda^2=\unit{-20}{dB}$ should be within experimental reach.

\ack

We thank D. Kienzler and T. Thiele for careful reading and comments on the manuscript, Matthias Troyer for useful help with numerical methods and Stefan Stahl for valuable electronics information. This work was supported by the Swiss NSF under grant no.~200021\_134776, and the NCCR QSIT.

\appendix
\section{Finite switching time - linear and sinusoidal potential displacements}\label{linsin}

Here we calculate and then compare the effect of two different transitions between the initial and transport potential wells for the case where voltage switching takes a finite time (see section \ref{finite}).

Let us consider first that the transitions from $t=0$ to $\tau$ and from $t=T$ to $T+\tau'$ are linear (as in figure \ref{fig:finiteswitch}), such that their slopes are $\dot{s}_\text{throw}=z_0/\tau$ and $\dot{s}_\text{catch}=z_0/\tau'$, respectively. In this case, equation \ref{eq:finswitch1} becomes
\be\label{eq:finswitch2}
\nonumber \alpha^\text{lin}(T+\tau')& = &i\sqrt{\frac{m}{2\hbar\omega}}\frac{z_0}{\tau}e^{-i\omega\tau'}f^\text{lin}(T),\\
 f^\text{lin}(T)& \approx & \omega\tau\left[e^{-i\omega T}\left(i-\frac{\omega\tau}{2}\right)+i-\frac{\omega\tau'}{2}\right],
\ee
where the approximation is for $\tau,\tau'\ll 1/\omega$. The value of $T$ which minimizes $\left|f^\text{lin}(T)\right|$ has the analytical form
\be\label{eq:finswitch4}
T_\text{min}^\text{lin}=-\frac{1}{\omega}\text{Im}\left\{\log\left(\frac{2i-\omega\tau}{-2i+\omega\tau'}\right)\right\},
\ee
and yields an $|\alpha^\text{lin}_\text{min}(T+\tau')|$ shown on figure \ref{fig:finswtchalpha}, for the specific case of a \caf ion at $\omega=2\pi\cdot(\unit{1}{\mega\hertz})$, $z_0=\unit{50}{\micro\meter}$ and $\tau=\unit{5}{\nano\second}$

A sinusoidal transition,
\be
\nonumber s_\text{throw}(t) & = & -\frac{z_0}{2}\left[1+\cos\left(\frac{\pi t}{\tau}\right)\right],\\
 s_\text{catch}(t) & = & \frac{z_0}{2}\left[1-\cos\left(\frac{\pi t}{\tau'}-\frac{\pi T}{\tau'}\right)\right],
\ee
resembles more closely the pulse shown in figure \ref{fig:pulse}, while still yielding an analytic solution:
%
\be\label{eq:finswitch3}
\nonumber \alpha^\text{cos}(T+\tau')& = & \sqrt{\frac{m\omega}{2\hbar}}\frac{\pi^2z_0}{2}e^{-i\omega\tau'}f^\text{cos}(T),\\
 f^\text{cos}(T) & \approx &\frac{2+i\omega\tau-\frac{(\omega\tau)^2}{2}}{\pi^2-(\omega\tau)^2}e^{-i\omega T}+\frac{2+i\omega\tau'-\frac{(\omega\tau')^2}{2}}{\pi^2-(\omega\tau')^2},
\ee
where the approximation is again for $\tau,\tau'\ll 1/\omega$. In this case,
\be
T_\text{min}^\text{cos}=-\frac{1}{\omega}\text{Im}\left\{\log\left(-\frac{\left(\pi^2-\omega^2\tau'^2\right)\left(-4-2i\omega\tau+\omega^2\tau^2\right)}{\left(\pi^2-\omega^2\tau^2\right)\left(-4-2i\omega\tau'+\omega^2\tau'^2\right)}\right)\right\}.
\ee

\begin{figure}
\centering
\resizebox{0.7\textwidth}{!}{
  \includegraphics{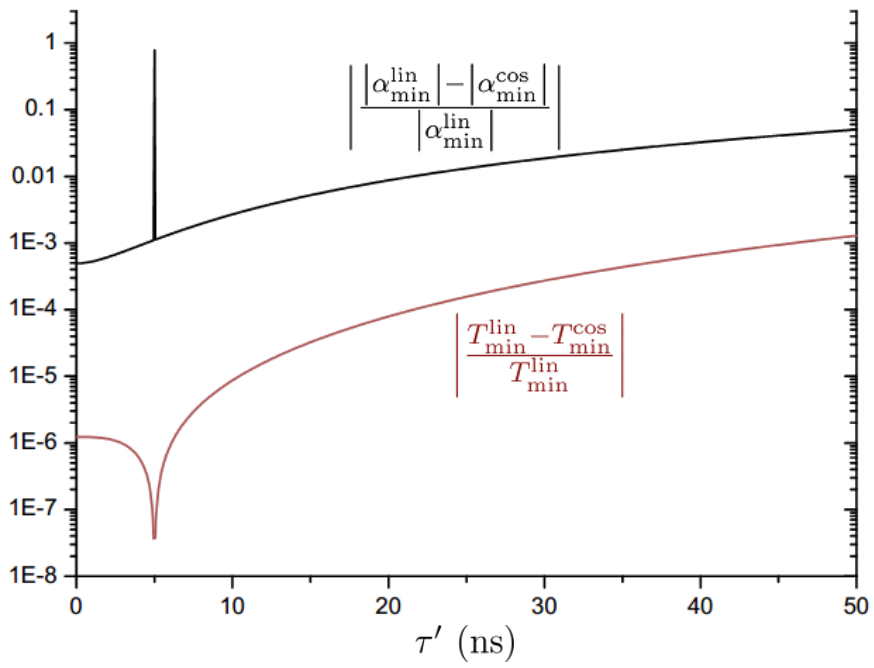}}
\caption{Relative difference between the linear and the cosine transitions for a \caf ion at $\omega=2\pi\cdot(\unit{1}{\mega\hertz})$, $z_0=\unit{50}{\micro\meter}$ and $\tau=\unit{5}{\nano\second}$ for $|\alpha_\text{min}|$ and $|T_\text{min}|$.}
\label{fig:finswtchalpha2}
\end{figure}

A comparison of the amplitude of the final coherent states for the two different examples (equations \ref{eq:finswitch2} and \ref{eq:finswitch3}) is shown on figure \ref{fig:finswtchalpha2}, for the same conditions as above. The small difference between them seems to indicate that the exact shape of $s_\text{catch,throw}(t)$ is not greatly important.

\section{Heating rate of squeezed vacuum states}\label{app}

Here we will calculate the lifetime of squeezed vacuum states, which can be generated in a trapped-ion experiment as discussed in section \ref{squeezing}. We will start by considering the transition rate out of an arbitrary state $\ket{\phi}$, given by
\be
\frac{d}{dt} \left\langle \left|\bra{\phi} \hat{U}(t) \ket{\phi}\right|^2\right\rangle,
\ee
where $\hat{U}(t)$ is the time-evolution operator and the right and left angle brackets average over all realizations.

We consider here the effect of a time-dependent perturbation Hamiltonian $\hat{H}(t)$ which can be written as
\be
\hat{H}(t) = -q E(t) \hat{z}, \label{eq:Hamiltonian}
\ee
where $q$ is the charge of the ion, $E(t)$ the electric field, and $\hat{z}$ is the coordinate of the ion center of mass. We restrict our study to a single dimension, since the normal modes of a trapped ion are independent and thus the extension to other dimensions is trivial. Heating (or decoherence) is measured as an average over many runs of an experiment, so $E(t)$ is a stochastic variable. If it is derived from a stationary process with fluctuations around a mean value of zero, we can write averaged quantities
\be
\langle E(t)\rangle = 0 \\
\langle E(t)E(t + \tau)\rangle = C(\tau),
\ee
where $C$ is a function to be determined experimentally, but which depends only on $\tau$.

A Hamiltonian of the form given in equation \ref{eq:Hamiltonian} gives evolution according to
\be
\hat{U}(t) = \hat{D}(\alpha_E(t)) e^{i \Phi(t)},
\ee
where $\hat{D}(\alpha_E)$ is the displacement operator by an amount $\alpha_E$ given by
\be
\alpha_E(t) = -\frac{i q a_0}{\hbar}\int_0^t \text{d}{t'}E(t')e^{i \omega t'}
\ee
in the interaction picture. Starting from an initial state $\ket{\phi}$, the resulting state is
\be
\hat{D}(\alpha_E(t)) e^{i \Phi(t)}\ket{\phi}.
\ee
The overlap with the initial state is
\be
e^{i \Phi(t)} \bra{\phi} \hat{D}(\alpha_E(t)) \ket{\phi},
\ee
for which the modulus squared gives the probability of overlap
\be
P_O(t) = \left|\bra{\phi} \hat{D}(\alpha_E(t)) \ket{\phi}\right|^2.
\ee

For a squeezed vacuum state
\be\label{eq:sqheating2}
\ket{\hat{\Sigma}(\xi)} = \hat{\Sigma}(\xi) \ket{0} \equiv e^{(\xi {\create}^2 - \xi^* {\destroy}^2)/2}\ket{0},
\ee
one can use \cite{99Metcalf}
\be
\nonumber \hat{\Sigma}(-\xi) \create \hat{\Sigma}(\xi) = \create \cosh{r} - \destroy e^{-i \Phi} \sinh{r}\\
\hat{\Sigma}(-\xi) \destroy \hat{\Sigma}(\xi) = \destroy \cosh{r} + \create e^{i \Phi} \sinh{r}
\ee
to find that
\be\label{eq:sqheating3}
P_\text{O}(t)=\text{exp}\left\{-|\alpha_E|^2 \frac{\cosh(2 r)}{2} - (\alpha_E^2 e^{i \Phi} + (\alpha_E^*)^2 e^{-i \Phi} )\frac{\sinh(2 r)}{2}\right\}.
\ee

The loss of overlap with the initial state will be the ensemble average of this quantity. In order to evaluate it, we need to find
\be\label{eq:expalphaEsq}
\left\langle e^{-|\alpha_E(t)|^2} \right\rangle,
\ee
with the exponent given by
\be\label{eq:alphaEsq2}
|\alpha_E(t)|^2 = \frac{q^2 a_0^2}{\hbar^2} \int_0^t dt' \int_0^t dt'' E(t') E(t'') e^{-i \omega_z (t'' - t')}.
\ee
We can relate the expectation value of the exponential in equation \ref{eq:expalphaEsq} to the exponential of the expectation value of its argument if we assume that the averaging time is short compared to the time scale over which the level populations vary, but large compared to the correlation time of the fluctuations. The expectation value of the argument can be found to be \cite{97Savard}
\be\label{eq:alphaEsq}
\left\langle |\alpha_E(t)|^2 \right\rangle = \frac{q^2 a_0^2}{2 \hbar^2}S_E(\omega_z) t,
\ee
where $S_E(\omega_z)$ is the spectral noise density in the trap (see equation \ref{eq:heatrate}).

This gives a decay constant (considering only the $|\alpha_E|^2$ term):
\be\label{eq:heatrate2}
\tau^{-1}=\frac{q^2}{4 m \omega \hbar}S_E(\omega_z)\frac{\cosh(2r)}{2}=\Gamma_{0\rightarrow 1}\frac{\cosh(2r)}{2},
\ee
with $\Gamma_{0\rightarrow 1}$ defined by equation \ref{eq:heatrate}.

The exponent in equation \ref{eq:sqheating3} also contains two terms with $\alpha_E^2$ and $(\alpha_E^*)^2$. The $\alpha_E^2$ term contains the integral
\be
\int_0^t dt' \int_0^t dt'' E(t') E(t'') e^{-i \omega_z (t'' + t')}.
\ee
This is a symmetric integral, in which we can make the substitution $\tau = t'' - t'$ to obtain
\be
\int_0^t dt' e^{-2 i \omega_z t'} \int_{-t'}^{t-t'} d\tau E(t') E(t' + \tau) e^{-i \omega_z \tau}.
\ee
Note the complex exponential in the integral over $t'$, not present in equation \ref{eq:alphaEsq2}. Under the same approximation as before, and again taking the ensemble average, we obtain an additional sinc function, meaning that this term tends to zero for $t\gg2 \pi/\omega$. This reflects the fact that the phase of the field in the Hamiltonian is uncertain and will be distributed over the full range from zero to $2 \pi$. Hence, it will be averaged to zero over many realizations or over long interaction times. Therefore, taking the ensemble average, we see that the loss of probability of overlap is at a rate $\tau^{-1}$ as given in equation \ref{eq:heatrate2}.


\section*{References}
\begin{thebibliography}{99}

\bibitem{67Dehmelt}
H.G. Dehmelt,  Adv. At. Mol. Phys. \textbf{3}, (1967) 53
\bibitem{96Ghosh}
P.K. Ghosh, \textit{Ion traps} (Clarendon, Oxford, 1996)
\bibitem{05Major}
F.G. Major, V.N. Gheorghe and G. Werth, \textit{Charged particle traps} (Springer, Berlin, 2005)
\bibitem{96Meekhof}
D.M. Meekhof, C. Monroe, B.E. King, W.M. Itano, and D.J. Wineland, Phys. Rev. Lett. \textbf{76} (1996) 1796
\bibitem{96Monroe}
C. Monroe, D.M. Meekhof, B.E. King, and D.J. Wineland, Science \textbf{272} (1996) 1131
\bibitem{00Myatt}
C.J. Myatt, B.E. King, Q.A. Turchette, C.A. Sackett, D. Kielpinski, W.M. Itano, C. Monroe, and D.J. Wineland, Nature \textbf{403}, (2000) 269
\bibitem{07McDonnell}
M.J. McDonnell, J.P. Home, D.M. Lucas, G. Imreh, B.C. Keitch, D.J. Szwer, N.R. Thomas, S.C. Webster, D.N. Stacey, and A.M. Steane, Phys. Rev. Lett. \textbf{98} (2007) 063603
\bibitem{05Haljan}
P.C. Haljan, K.-A. Brickman, L. Deslauriers, P.J. Lee, and C. Monroe, Phys. Rev. Lett. \textbf{94} (2005) 153602
\bibitem{10Zahringer}
F. Z\"ahringer, G. Kirchmair, R. Gerritsma, E. Solano, R. Blatt, and C.F. Roos, Phys. Rev. Lett. \textbf{104} (2010) 100503
\bibitem{03Leibfried}
D. Leibfried, B. DeMarco, V. Meyer, M. Rowe, A. Ben-Kish, J. Britton, W.M. Itano, B.J.
Jelenkovi\'c, C. Langer, T. Rosenband, and D.J. Wineland, Phys. Rev. Lett. \textbf{89} (2002) 247901
\bibitem{03SchmidtKaler}
F. Schmidt-Kaler, H. H\"affner, M. Riebe, S. Gulde, G. Lancaster, T. Deuschle, C. Becher, C. Roos, J. Eschner, R. Blatt, Nature \textbf{422} (2003) 408
\bibitem{09Jost}
J.D. Jost, J.P. Home, J. M. Amini, D. Hanneke, R. Ozeri, C. Langer, J.J. Bollinger, D. Leibfried, and D.J. Wineland, Nature (London) \textbf{459} (2009) 683
\bibitem{11Brown}
K.R. Brown,	C. Ospelkaus, Y. Colombe, A.C. Wilson, D. Leibfried, and D.J. Wineland, Nature \textbf{471}, (2011) 196
\bibitem{11Harlander}
M. Harlander, R. Lechner, M. Brownnutt, R. Blatt, and W. H\"ansel, Nature \textbf{471}, (2011) 200
\bibitem{06Schulz}
S. Schulz, U. Poschinger, K. Singer, and F. Schmidt-Kaler, Progress of Physics, Wiley \textbf{54, No. 8-10} (2006) 648
\bibitem{03Steane}
A.M. Steane, Phys. Rev. A \textbf{68} (2003) 042322
\bibitem{98Wineland}
D.J. Wineland, C. Monroe, W.M. Itano, D. Leibfried, B.E. King and D.M. Meekhof,  J. Res. Natl. Inst. Stand. Technol. \textbf{103}, (1998) 259
\bibitem{02Kielpinski}
D. Kielpinski, C.R. Monroe, and D.J. Wineland, Nature \textbf{417} (2002) 709
\bibitem{02Rowe}
M.A. Rowe, A. Ben-Kish, B. DeMarco, D. Leibfried, V. Meyer, J. Beall, J. Britton, J. Hughes, W.M. Itano, B. Jelenkovic, C. Langer, T. Rosenband, and D.J. Wineland, Quantum Information and Computation \textbf{2}, (2002) 257-271
\bibitem{12Bowler}
R. Bowler, J. Gaebler, Y. Lin, T.R. Tan, D. Hanneke, J.D. Jost, J.P. Home, D. Leibfried, and D.J. Wineland, Phys. Rev. Lett. \textbf{109}, (2012) 080502
\bibitem{12Walther}
A. Walther, F. Ziesel, T. Ruster, S.T. Dawkins, K. Ott, M. Hettrich, K. Singer, F. Schmidt-Kaler, and U. Poschinger, Phys. Rev. Lett. \textbf{109}, (2012) 080501
\bibitem{06Hensinger}
W.K. Hensinger, S. Olmschenk, D. Stick, D. Hucul, M. Yeo, M. Acton, L. Deslauriers, C. Monroe, and J. Rabchuk, Appl.
Phys. Lett. \textbf{88} (2006) 034101
\bibitem{10Amini}
J.M. Amini, H. Uys, J.H. Wesenberg, S. Seidelin, J. Britton, J.J. Bollinger, D. Leibfried, C. Ospelkaus, A.P. VanDevender, and D.J.
Wineland, New J. Phys. \textbf{12} (2010) 033031
\bibitem{11Moehring}
D.L. Moehring, C. Highstrete, M.G. Blain, K. Fortier, R. Halti, D. Stick, and C. Tigges, New J. Phys. \textbf{13} (2011) 075018
\bibitem{09Blakestad}
R.B. Blakestad, C. Ospelkaus, A.P. VanDevender, J.M. Amini, J. Britton, D. Leibfried, and D.J. Wineland, Phys. Rev. Lett. \textbf{102} (2009) 153002
\bibitem{03Barrett}
M.D. Barrett, B. DeMarco, T. Schaetz, V. Meyer, D. Leibfried, J. Britton, J. Chiaverini, W.M. Itano, B. Jelenkovic, J.D. Jost, C. Langer, T. Rosenband, and D.J. Wineland, Phys. Rev. A \textbf{68} (2003) 042302
\bibitem{09Home}
J.P. Home, D. Hanneke, J.D. Jost, J.M. Amini, D. Leibfried, and D.J. Wineland, Science \textbf{325} (2009) 1227
\bibitem{09Home2}
J.P. Home, M.J. McDonnell, D.J. Szwer, B.C. Keitch, D.M. Lucas, D.N. Stacey, and A.M. Steane, Phys. Rev. A \textbf{79} (2009) 050305(R)
\bibitem{10Hanneke}
D. Hanneke, J.P. Home, J.D. Jost, J. Amini, D. Leibfried, and D.J. Wineland, Nature Physics 6, \textbf{6} (2010) 13
\bibitem{09Serafini}
A. Serafini, A. Retzker, and M.B. Plenio, New J. Phys. \textbf{11} (2009) 023007
\bibitem{09Serafini2}
A. Serafini, A. Retzker, and M.B. Plenio, Quantum Inf. Proc. \textbf{8} (2009) 619
\bibitem{00Turchette}
Q.A. Turchette, D. Kielpinski, B.E. King, D. Leibfried, D.M. Meekhof, C.J. Myatt, M.A. Rowe, C.A. Sackett, C.S. Wood, W.M. Itano, C. Monroe, and D.J. Wineland, Phys. Rev. A \textbf{61} (2000) 063418
\bibitem{12BowlerPrivComm}
R. Bowler, \emph{Private communication}
\bibitem{12Alonso}
J. Alonso, R. Habl\"utzel, T. Kartanas, B. Keitch, F. Leupold, S. Stahl and J. Home, \emph{Publication in preparation}
\bibitem{74HC4066M}
Texas Instruments, datasheet from CD54HC4066, CD74HC4066, CD74HCT4066 High-Speed CMOS Logic Quad Bilateral Switch
\bibitem{filtercs}
See e.g.: J. Labaziewicz, PhD thesis (Harvard, 2008); S.A. Schulz, PhD thesis (Ulm, 2009); reference \cite{10Blakestad}
\bibitem{05Chiaverini}
J. Chiaverini, R.B. Blakestad, J. Britton, J.D. Jost, C. Langer, D. Leibfried, R. Ozeri, and D.J. Wineland, Quantum Inf. Comput. \textbf{5} (2005) 419.
\bibitem{09Amini}
J.M. Amini, J. Britton, D. Leibfried, and D.J. Wineland, \textit{Atom Chips} (Wiley-VCH, New York, 2009)
\bibitem{09Antohi}
P.B. Antohi, D. Schuster, G.M. Akselrod, J. Labaziewicz, Y. Ge, Z. Lin, W.S. Bakr, and I. L. Chuang, Rev. Sci. Inst. \textbf{80} (2009) 013103
\bibitem{02Wineland}
D.J. Wineland, Int. School of Physics Enrico Fermi, F. De Martini and C. Monroe, eds., IOS Press, Amsterdam \textbf{148} (2002) 165
\bibitem{08Couvert}
A. Couvert, T. Kawalec, G. Reinaudi, and D. Guery-Odelin, Europhys. Lett. \textbf{83} (2008) 13001
\bibitem{08Wesenberg}
J.H. Wesenberg, Phys. Rev. A \textbf{78} (2008) 063410
\bibitem{09Schmied}
R. Schmied, J.H. Wesenberg, and D. Leibfried, Phys. Rev. Lett. \textbf{102} (2009) 233002
\bibitem{10Schmied}
R. Schmied, New J. Phys. \textbf{12} (2010) 023038
\bibitem{10Blakestad}
R.B. Blakestad, PhD thesis (University of Colorado, 2010)
\bibitem{06Home}
J.P. Home, and A. Steane, Quantum Information and Computation \textbf{6} (2006) 289
\bibitem{93Suzuki}
M. Suzuki, Physica A \textbf{194} (1993) 432
\bibitem{picosec}
See e.g. Berkeley Nucleonics, datasheet from Model 745 \unit{250}{\femto\second} Digital Delay Generator; Stanford Research Systems, datasheet from Model DG535 Digital Delay/Pulse Generator; Picosecond Pulse Labs, datasheet from Model 12000 165MHz Digital Pulse/Pattern Generator;  A. Metzger, C.E. Chang, P.M. Asbeck, K.C. Wang, K. Pedrotti, A. Price, A. Campana, D. Wu, J. Liu, and S. Beccue, Gallium Arsenide Integrated Circuit (GaAs IC) Symposium, Technical Digest 1997, 19th Annual (1997) 109-112,
\bibitem{11Lau}
H.-K. Lau, and D.F.V. James, Phys. Rev. A \textbf{83} (2011) 062330
\bibitem{98Steane}
A. Steane, Rep. Prog. Phys. \textbf{61} (1998) 117
\bibitem{98James}
D.F.V . Phys. Rev. Lett. \textbf{81} (1998) 317
\bibitem{12Schabinger}
B. Schabinger, S. Sturm, A. Wagner, J. Alonso, W. Quint, G. Werth, and K. Blaum, Eur. Phys. J. D \textbf{66} (2012) 71
\bibitem{88Henry}
R.W. Henry and S.C. Glotzer, Amer. J. Phys. \textbf{56} (1988) 318-328
\bibitem{10Cormick}
C. Cormick, and J.P. Paz,  Phys. Rev. A \textbf{81} (2010) 022306
\bibitem{03Eisert}
J. Eisert and M.B. Plenio, Int. J. Quant. Inf. \textbf{1} (2003) 479
\bibitem{03Wolf}
M.M. Wolf, J. Eisert, and M.B. Plenio,  Phys. Rev. Lett. \textbf{90} (2003) 047904
\bibitem{97Barnett}
S.M. Barnett and P.M. Radmore, \textit{Methods in theoretical quantum optics} (Clarendon, Oxford, 1997)
\bibitem{11Hite}
D.A. Hite, Y. Colombe, A.C. Wilson, K.R. Brown, U. Warring, R. Joerdens, J.D. Jost, D.P. Pappas, D. Leibfried, and D.J. Wineland, arXiv:1112.5419 (2011)
\bibitem{11Allcock}
D.T.C. Allcock, L. Guidoni, T.P. Harty, C.J. Ballance, M.G. Blain, A.M. Steane, and D.M. Lucas, New J. Phys. \textbf{13} (2011) 123023
\bibitem{08Labaziewicz}
J. Labaziewicz, Y. Ge, D. R. Leibrandt, S.X. Wang, R. Shewmon, and I.L. Chuang, Phys. Rev. Lett. \textbf{101} (2008) 180602
\bibitem{08Vahlbruch}
H. Vahlbruch, M. Mehmet, S. Chelkowski, B. Hage, A. Franzen, N. Lastzka, S. Go\ss ler, K. Danzmann, and R. Schnabel, Phys. Rev. Lett. \textbf{100} (2008) 033602
\bibitem{97Scully}
M.O. Scully and M.S. Zubaira, \textit{Quantum Optics} (Cambridge University Press, Cambridge, 1997)
\bibitem{97Breitenbach}
G. Breitenbach, S.Schiller, and J. Mlynek, Nature \textbf{387}, (1997) 471
\bibitem{05Braunstein}
S.L. Braunstein and P. van Loock, Rev. Mod. Phys. \textbf{77} (2005) 513-577
\bibitem{04Eisert}
J. Eisert, M.B. Plenio, S. Bose, and J. Hartley, Phys. Rev. Lett. \textbf{93} (2004) 190402
\bibitem{04Tian}
L. Tian, P. Rabl, R. Blatt, and P. Zoller, Phys. Rev. Lett. \textbf{92} (2004) 247902
\bibitem{99Metcalf}
H.J. Metcalf and P. van der Straten, \textit{Laser cooling and trapping} (Springer-Verlag, New York, 1999)
\bibitem{97Savard}
T.A. Savard, K.M. O'Hara, and J.E. Thomas, Phys. Rev. A \textbf{56} (1997) R1095



\end{thebibliography}
\end{document}